\pdfoutput=1
\documentclass[sn-mathphys]{sn-jnl}

\jyear{2022}%

\usepackage{subfigure}
\usepackage{multirow}
\usepackage{epstopdf}
\usepackage{enumerate}
\catcode`\|=12
\theoremstyle{thmstyleone}%
\theoremstyle{thmstyletwo}%
\theoremstyle{thmstylethree}%
\begin{document}

\title{Quantum Symbolic Execution}
\author[1,2]{Jiang Nan}
\author[1]{Wang Zichen}
\author*[3,4]{Wang Jian}\email{wangjian@bjtu.edu.cn}
\affil[1]{The Faculty of Information Technology, Beijing University of Technology, Beijing 100124, China}
\affil[2]{Beijing Key Laboratory of Trusted Computing, Beijing 100124, China}
\affil[3]{School of Computer and Information Technology, Beijing Jiaotong University, Beijing
	100044, China}
\affil[4]{Beijing Key Laboratory of Security and Privacy in Intelligent Transportation, Beijing Jiaotong University, Beijing 100044, China}

\abstract{With advances in quantum computing, researchers can now write and run many quantum programs.
However, there is still a lack of effective methods for debugging quantum programs. In this paper, quantum symbolic execution (QSE) is proposed to generate test cases, which helps to finding bugs in quantum programs. The main idea of quantum symbolic execution is to find the suitable test cases from all possible ones (i.e. test case space). It is different from the way of classical symbol execution, which gets test cases by calculating instead of searching. QSE utilizes quantum superposition and parallelism to store the test case space with only a few qubits. According to the conditional statements in the debugged program, the test case space is continuously divided into subsets, subsubsets and so on. Elements in the same subset are suitable test cases that can test the corresponding branch in the code to be tested. QSE not only provides a possible way to debug quantum programs, but also avoids the difficult problem of solving constraints in classical symbolic execution.}


\maketitle

\section{Introduction}\label{sec1}
Quantum computing has attracted much attention, because quantum superposition, entanglement and other properties can greatly improve the efficiency of computing \cite{nielsen2002quantum,jiang2021programmable}. In recent years, with the development of quantum computer hardware \cite{zhong2020quantum,arute2019quantum}, quantum software and quantum programming  \cite{broughton2020tensorflow,cross2018ibm,paolini2019qpcf,selinger2004towards} has also been greatly developed. Researchers can write and run many quantum algorithms that have been proposed before but cannot be implemented due to limitations, such as Grover's algorithm \cite{adedoyin2018quantum}, quantum principal component analysis algorithm \cite{he2020exact}, quantum phase estimation \cite{o2019quantum}, and ${etc}$. 
In the process of writing quantum programs, some errors will inevitably occur \cite{paltenghi2021bugs,wang2018quanfuzz,miranskyy2020your}. 
For example, Zhao \cite{zhao2021identifying} defined a few bugs that focus on misuses of features of the quantum programming language --- Qiskit \cite{cross2018ibm}.
Huang \cite{huang2019statistical} also recorded some bugs in the Scaffold compiler \cite{javadiabhari2015scaffcc}. 
For quantum programs we still need to take corresponding measures to find these errors and fix them. 
Due to the characteristics of quantum computing, we cannot debug programs as in the classical environment. 
This difficulty in debugging quantum programs hinders the development of quantum computing. 
An effective quantum program debugging scheme is needed.

Researchers have proposed some methods for debugging quantum programs, including
quantum unit tests \cite{bright2017microsoft}, quantum assertions \cite{huang2019statistical,liu2020quantum,li2019proq,liu2021systematic}, and ${etc}$.
Unit tests are used to determine whether a specific function is correct under a specific condition.
The role of the assertion is that when the program executes to the assertion, the corresponding assertion should be true, and if the assertion is not true, the program should terminate execution.
These methods have corresponding quantum versions. 
However, these methods are not very good to meet the needs.
Currently, assertions in the quantum environment include statistical assertions \cite{huang2019statistical} based on classical observations, dynamic runtime assertions \cite{liu2020quantum} that use auxiliary qubits to obtain information indirectly, a projection-based runtime assertion \cite{li2019proq}, and dynamic assertion \cite{liu2021systematic} that extend dynamic runtime assertions \cite{liu2020quantum}.
These assertions have two main shortcomings.
Firstly, they are mostly used when an error has occurred during the running of the program or when the programmer suspects that there is an error somewhere in the program. 
Just like people do not directly set breakpoints on the entire program, but often set breakpoints only when the output is not as expected.
Secondly, the use of assertions relies on the prediction of results. 
They need to compare the actual output with the expected result to judge whether the program is error. 
This is not simple for quantum programs. 
Microsoft's ${Q\#}$ \cite{bright2017microsoft} provides a method for unit testing of quantum programs, which tests a unit of a quantum program individually to verify whether it meets expectations, and internally still uses assertions to achieve this goal.
There is another method Quito (quantum input output coverage) \cite{ali2021assessing}.
The biggest contribution of this paper is to define three coverage criteria for the input and output of quantum program debugging.
But the biggest flaw of this method is that it still uses statistical analysis to determine test pass and fail, which certainly does not reduce the complexity of quantum program debugging.
Therefore, they cannot meet the programmer's needs for quantum program debugging very well. 

Only unit tests and assertion cannot meet the needs of program debugging. In classical program debugging field, symbolic execution is another important debug method and it has appeared much earlier \cite{king1976symbolic}. With the development of constraint solving technology, symbolic execution has become an effective technology for generating high-coverage test cases \cite{cadar2013symbolic} and been widely used in different areas such as software testing, analysis and verification \cite{zhao40smart,yang2019cache,wang2017cached}.

This paper proposes a quantum symbolic execution (QSE) method, which focuses on generating high-coverage test cases for quantum programs. QSE uses quantum superposition and parallel characteristics to store the test case space with only a few qubits. According to the conditional statements in the debugged program, the test case space is continuously divided into subsets. Elements in the same subset are suitable test cases that can test the corresponding branch in the code to be tested. QSE not only provides a possible way to debug quantum programs, but also avoids the difficult problem of solving constraints in classical symbolic execution. 

\section{Related Works}
In this section, we briefly introduce the classical symbolic execution and some existing quantum modules that will be used in QSE.
\subsection{classical symbolic execution (CSE)} \label{sec21}
Programs often have conditional statements, and each branch represents an execution path to the program. In software testing, symbolic execution is a way to generate test cases that cover each execution path. Symbolic execution works by two steps:
\begin{enumerate}[(1)]
	\item creating execution paths, and
	\item using a constraint solver to calculate the answers to the execution paths, i.e., generating test cases.
\end{enumerate}

To formally accomplish this task, symbolic execution maintains two states globally: a symbolic state ${\sigma}$, which maps variables to symbolic expressions, and symbolic path constraints ${PC}$s, which are quantifier-free first-order logical formulas over symbolic expressions. 
At the beginning of a symbolic execution, ${\sigma}$ is initialized to an empty map and ${PC}$ is initialized to ${true}$. Both ${\sigma}$ and ${PC}$ are populated during the course of symbolic execution. The update rule of ${\sigma}$ is:
\begin{enumerate}
	\item[$\bullet$] At every read statement ${var=sym\_input()}$ that receives program input, symbolic execution adds the mapping ${var \mapsto s}$ to ${\sigma}$, where ${s}$ is a fresh symbolic value. 
	\item[$\bullet$] At every assignment ${v=e}$, symbolic execution updates ${\sigma}$ by mapping ${v}$ to ${\sigma(e)}$, where ${\sigma(e)}$ is the mapping of the symbolic state ${\sigma}$ to the expression ${e}$.
\end{enumerate}
The update rule of ${PC}$ is:
\begin{enumerate}
	\item[$\bullet$] At every conditional statement ${if \; (e) \; S1 \; else \; S2}$, ${PC}$ is updated to
	${PC_1=PC\land \sigma(e)}$ (``then'' branch) and ${PC_2={PC} \land \lnot \sigma(e)}$ (``else'' branch). 
\end{enumerate}

\begin{figure}[htbp]
	\centering
	\includegraphics[scale=0.3]{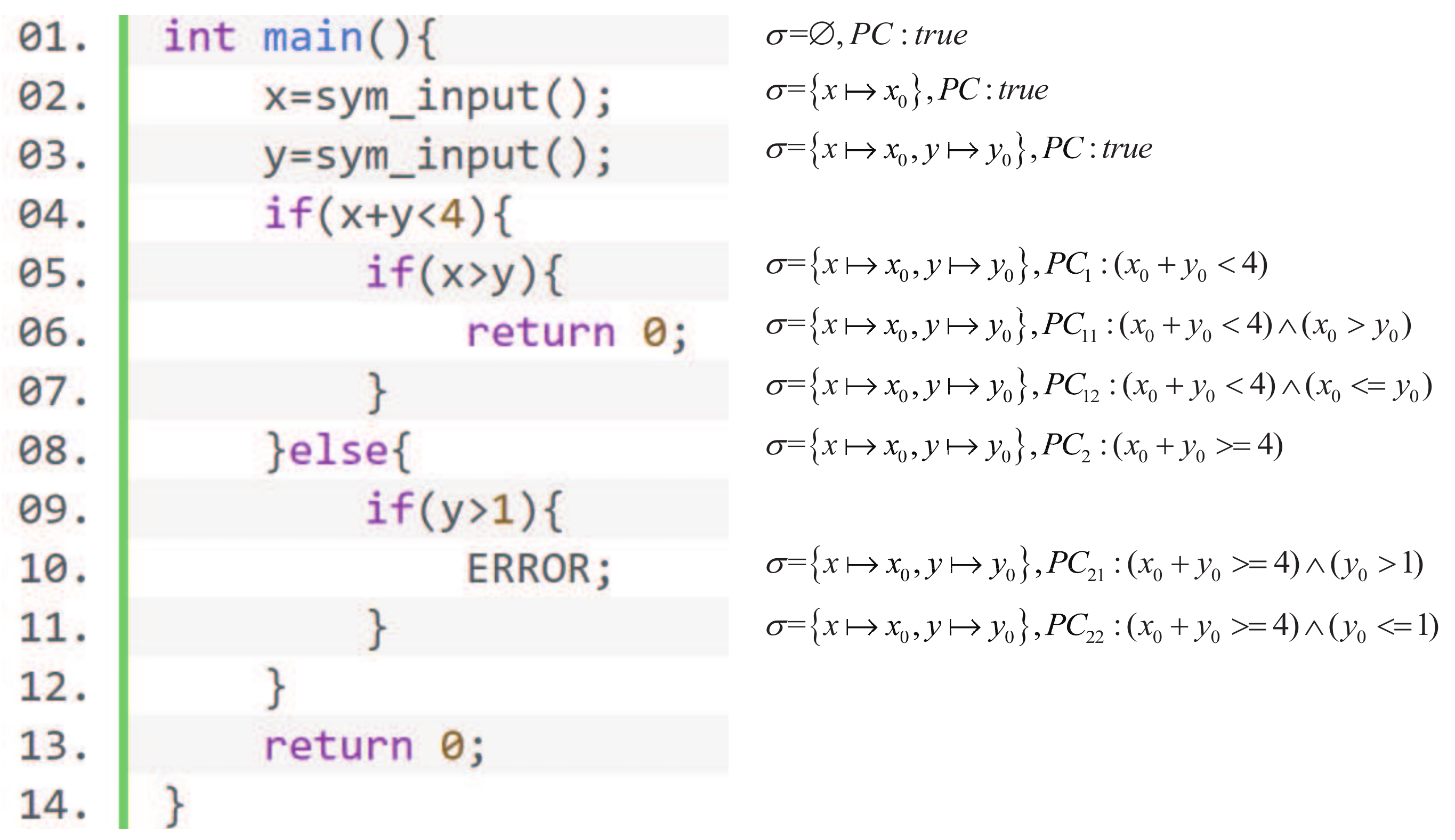}
	\caption{An example to illustrate symbolic execution}
	\label{fig1}
\end{figure}

For example, the symbolic execution of the code in Fig. \ref{fig1} starts with an empty symbolic state $\sigma$ and a symbolic path constraint ${true}$. After Line 03, ${\sigma=\{x \mapsto x_0, y \mapsto y_0\}}$; after Line 05, a path constraint ${(x_0+y_0<4)\land(x_0>y_0)}$ is created; and after Line 09, a path constraint ${(x_0+y_0\geq4)\land(y_0>1)}$ is created. Finally, there are 4 path constrains: $PC_{11}$, $PC_{12}$, $PC_{21}$, and $PC_{22}$. Each path constraint is solved with a constraint solver to obtain test cases. $\{x=2,y=1\}$, $\{x=1,y=2\}$, $\{x=3,y=2\}$, and $\{x=4,y=1\}$ are the possible outputs of the constraint solver for $PC_{11}$, $PC_{12}$, $PC_{21}$, and $PC_{22}$ respectively, i.e., they are suitable test cases.

\begin{figure}[htbp]
	\centering
	\includegraphics[scale=0.8]{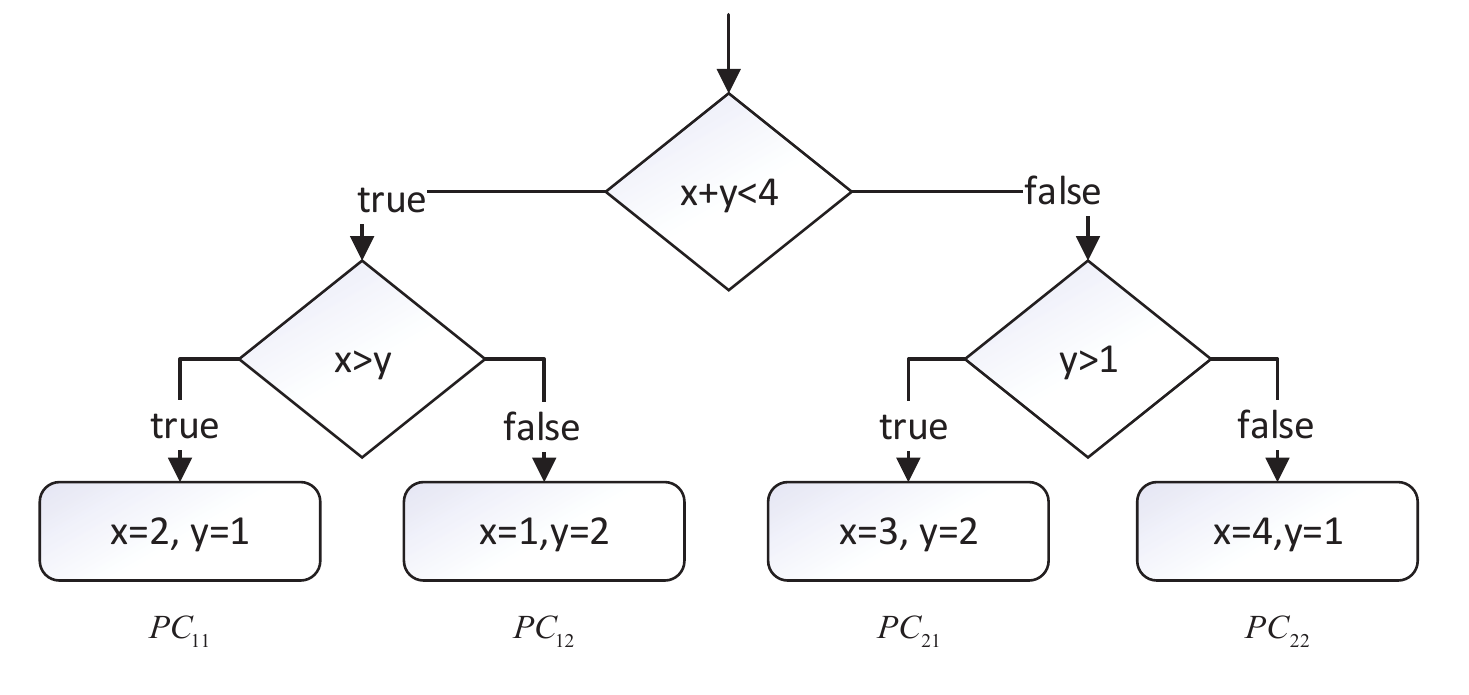}
	\caption{The execution tree for the example in Fig. \ref{fig1}}
	\label{fig2}
\end{figure}

All the execution paths of a program can be represented using a tree, called the execution tree. 
For example, Fig. \ref{fig2} gives the execution tree of the code in Fig. \ref{fig1}.  The 4 branches correspond to the 4 path constrains.

\subsection{related quantum modules}\label{sec24}
\begin{figure}[htbp]
	\centering
	\subfigure[quantum adder]{
		\includegraphics[scale=1]{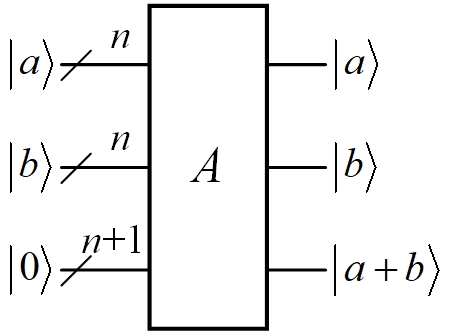}\label{fig3}
	}
	\quad
	\subfigure[quantum multiplier]{
		\includegraphics[scale=1]{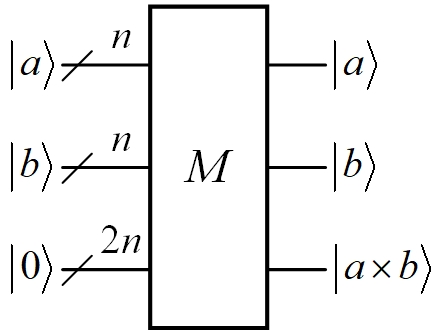}\label{fig4}
	}
	\quad
	\subfigure[quantum comparator]{
		\includegraphics[scale=1]{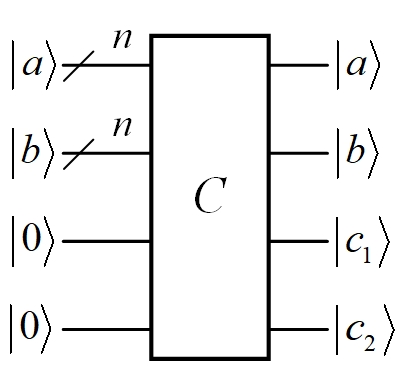}\label{fig5}
	}
	\caption{Three quantum modules}
\end{figure}

Suppose ${a}$ and ${b}$ are two ${n}$-qubit binary numbers, quantum adder \cite{chang2019design} ``${A}$'' implements addition of two qubits:
\begin{equation*}
	A(|ab \rangle |0 \rangle ^{\otimes n+1})= |ab \rangle |a+b \rangle.
\end{equation*}
The quantum module is shown in Fig. \ref{fig3}.

Quantum multiplier \cite{2019Quantum} ``${M}$'' implements multiplication of two qubits:
\begin{equation*}
	M(|ab\rangle |0\rangle^{\otimes 2n})= |ab\rangle |a\times b\rangle.
\end{equation*}
The quantum module is shown in Fig. \ref{fig4}.

The quantum comparator \cite{wang2012design} ``${C}$'' is used to compare two binary numbers. ${c_1}$ and ${c_2}$ are two 1-qubit outputs to record the comparison:
\begin{equation*}
	C(|ab\rangle|00\rangle)=|ab\rangle|c_1c_2\rangle.
\end{equation*}
When ${a>b}$, ${\left|c_1c_2\right\rangle=\left|10\right\rangle}$; when ${a<b}$, ${\left|c_1c_2\right\rangle=\left|01\right\rangle}$; and when ${a=b}$, ${\left|c_1c_2\right\rangle=\left|00\right\rangle}$. The module is shown in Fig. \ref{fig5}.

\section{Quantum symbolic execution}
In this section, we first give the workflow of quantum symbolic execution. 
Then we explain how to prepare the initial test case space and use relational operators, logical operators to delineate subspaces. 
Then we give the overall framework of QSE.
Finally give an example to illustrate.

\subsection{main idea} \label{sec31}
In Section \ref{sec21}, we briefly describe the process of symbolic execution in the classical environment. Generally speaking, it first traverses the program to collect the path constraints, and then uses the constraint solver to calculate a set of inputs that meet the path constraints.

Quantum symbolic execution is completely different, which works by two steps:
\begin{enumerate}[(1)]
	\item generating a test case space that includes all possible test cases, and
	\item according to the conditional statements in the code to be tested, partitioning the test case space into subspaces, and each subspace contains all the test cases that fit into a path constraint.
\end{enumerate}

\begin{figure}[htbp]
	\centering
	\includegraphics[scale=0.8]{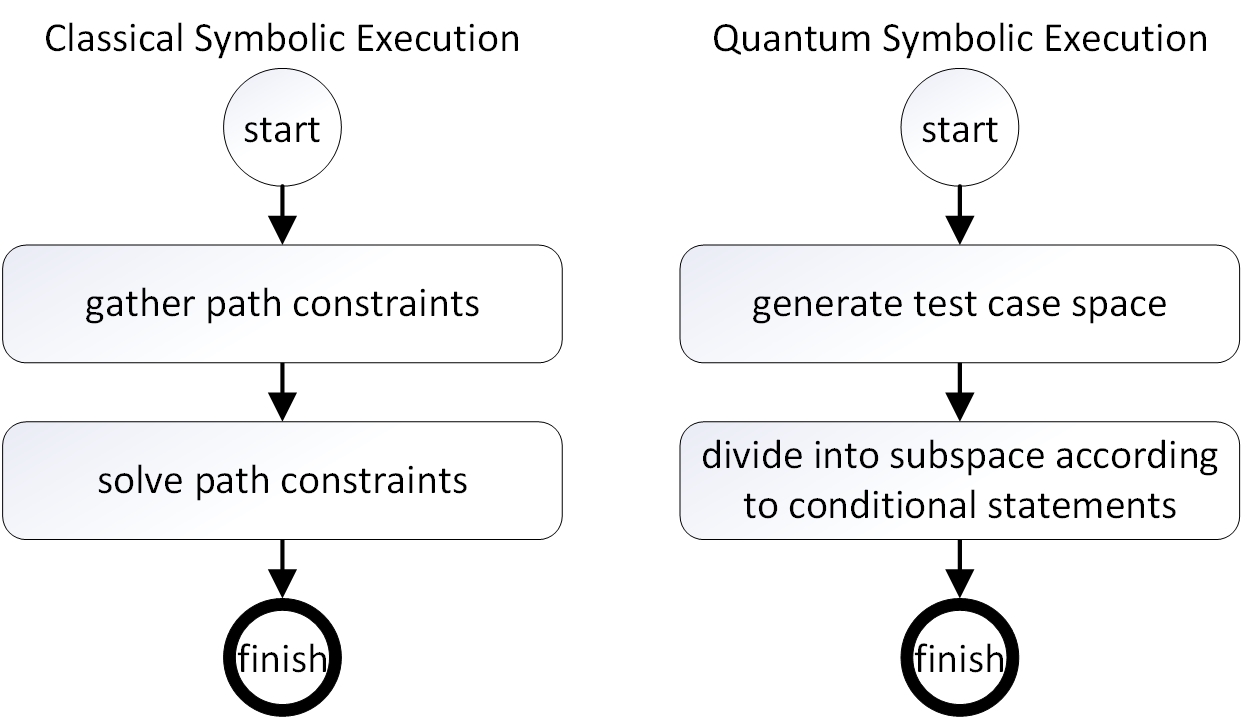}
	\caption{The contrast between classical symbolic execution and quantum symbolic execution.}
	\label{fig16}
\end{figure}

Fig. \ref{fig16} contrasts classical symbolic execution and quantum symbolic execution.

QSE uses two quantum registers: $|s\rangle$ and $|c\rangle$, where 
\begin{equation}
	|q\rangle=\frac{1}{\sqrt{2}^n}\sum_{i=0}^{2^{n}-1}|s_i\rangle\otimes|c_i\rangle
\end{equation}
$|s\rangle=|s^{n-1}s^{n-2}\cdots s^0\rangle$ consists of $n$ qubits and $s_i$ is a value used to represent a test case. $|c\rangle=|c^{m-1}c^{m-2}\cdots c^0\rangle$ consists of $m$ qubits and is the flag to subspace. $|s\rangle$ and $|c\rangle$ entangle together to realize the partition of $|s\rangle$: $s_i$ with the same $c_i$ belongs to the same subset, i.e. test cases for the same branch. $|s\rangle$ and $|c\rangle$ are collectively referred to as $|q\rangle$.

The flag $|c\rangle$ plays an important role in QSE, and it is gradually modified as the conditional statements in the code to be tested. Different conditions correspond to different ways to modify $|c\rangle$. Therefore, it is necessary to know how many types of conditions there are when programming. According to \cite{prata2014c, bruce2006java, eric2015python}, the conditions mainly include relational operation in Table \ref{table1} and logical operation in Table \ref{table2}.
\begin{table}
	\caption{relational operation}
	\begin{center}
		\begin{tabular}{cc} 
			\hline
			relational operators & meaning  \\
			\hline
			${<}$ & less than   \\
			${<=}$ & less than or equal to \\
			${>}$ & greater than  \\
			${>=}$ & greater than or equal to \\
			${==}$ & equal to  \\
			${!=}$ & not equal to  \\
			\hline
		\end{tabular} \label{table1}
	\end{center}
\end{table}

\begin{table}
	\caption{logical operation}
	\begin{center}
		\begin{tabular}{cc} 
			\hline
			logical operators & meaning  \\
			\hline
			${\&\&}$ & AND   \\
			${||}$ & OR  \\
			${!}$ &  NOT  \\
			\hline
		\end{tabular} \label{table2}
	\end{center}
\end{table}

The effects of relational and logical operations on $|c\rangle$ will be described in detail in Sections \ref{relationaloperator} and \ref{logicaloperator}, respectively.

\subsection{Preparation of the test case space}
Prepare $m+n$ qubits and set all of them to $|0\rangle$. The initial state of $|q\rangle$ is
\begin{equation}
	|q\rangle_0 =|0\rangle^{\otimes n}\otimes|0\rangle^{\otimes m}
\end{equation}
i.e., $|s\rangle_0=|0\rangle^{\otimes n}$ and $|c\rangle_0=|0\rangle^{\otimes m}$.

$n$ $H$ quantum gates and $m$ $I$ quantum gates are used to transform the initial state $|q\rangle_0$ to state $|q\rangle_1$, where
$$
H=\frac{1}{\sqrt{2}}\left[
\begin{array}{cc}
	1 & 1 \\
	1 & -1 
\end{array}
\right],\ \ I=\left[
\begin{array}{cc}
	1 & 0 \\
	0 & 1 
\end{array}
\right]
$$

The quantum preparation of the test case space can be expressed as $U_1$:
\begin{equation}
	U_1 = H^{\otimes n}\otimes I^{\otimes m}
\end{equation}

\begin{figure}[htbp]
	\centering
	\includegraphics[scale=1]{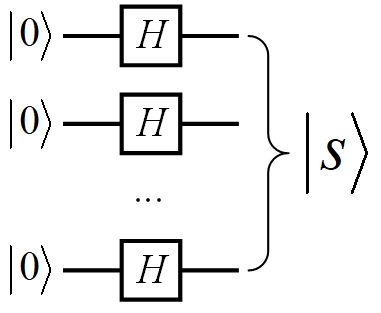}
	\caption{The preparation of the test case space}
	\label{fig6}
\end{figure}

$U_1$ changes the initial state $|q\rangle_0$ to the test case space:
\begin{equation}\label{U1}
	\begin{split}
		|q\rangle_1=&U_1(|q\rangle_0)\\
		=&H^{\otimes n}(|s\rangle_0)\otimes I^{\otimes m}(|c\rangle_0)\\
		=&(H|0\rangle)^{\otimes n}\otimes (I|0\rangle)^{\otimes m}\\
		=&\frac{1}{\sqrt 2}(|0\rangle+|1\rangle) \otimes \frac{1}{\sqrt 2}(|0\rangle+|1\rangle) \otimes  \cdots  \otimes \frac{1}{\sqrt 2 }(|0\rangle+|1\rangle) \otimes|0\rangle^m\\
		=&\frac{1}{{\sqrt 2 }^n}\left( |0 \cdots 00\rangle  + |0 \cdots 01\rangle +\cdots+|1 \cdots 11\rangle\right)\otimes|0\rangle^m\\
		=&\frac{1}{{\sqrt 2 }^n}\left( |0\rangle  + |1\rangle+\cdots +|2^n-1\rangle\right)\otimes|0\rangle^m\\
		=&\frac{1}{\sqrt{2}^n}\sum_{i=0}^{2^n-1}|i\rangle\otimes|0\rangle^m\\
		=&|s\rangle\otimes|0\rangle^m
	\end{split}
\end{equation}
where $|s\rangle=\frac{1}{\sqrt{2}^n}\sum_{i=0}^{2^n-1}|i\rangle$. The quantum circuit is shown in Fig. \ref{fig6}.

Eq. (\ref{U1}) shows that the test case space  $|s\rangle$ stores all integers from 0 to $2^n-1$, which are all the possible test cases. If the code to be tested contains $l$ ($l>1$) variables $x_1,x_2,\cdots,x_l$, $|s\rangle$ is still able to store all possible test cases. Divide the $n$ qubits of $|s\rangle$ into $l$ parts and each part stores all the possible value of a variable. The $i$th part contains $n_i$ qubits $|s_{x_i}\rangle=|s_{x_i}^{n_i-1}s_{x_i}^{n_i-2}\cdots s_{x_i}^{0}\rangle$, where $n=\sum_{i=1}^{l}n_i$. For example, the code in Fig. \ref{fig1} has two variables: $x$ and $y$. They contain 3 and 2 qubits respectively. Hence,
$$
|s\rangle=|s_xs_y\rangle=\frac{1}{\sqrt{2}^3}\sum_{i=0}^{7}|i\rangle\otimes\frac{1}{\sqrt{2}^2}\sum_{i=0}^{3}|i\rangle=\frac{1}{\sqrt{2}^5}\sum_{i=0}^{31}|i\rangle
$$

\subsection{Relational operator} \label{relationaloperator}
Relational operators compare two numbers. Therefore, QSE uses the quantum comparator to divide the test case space. Section 2.2 shows that the quantum comparator has two output qubits: $|c_1c_2\rangle$. Suppose they correspond to some two adjacent qubits in $|c\rangle=|c^{m-1}c^{m-2}\cdots c^0\rangle$, and mark them as $|c^{i}c^{i-1}\rangle$. Combining Table \ref{table1}, we can get the relationship between the relational operators and the state of the output qubits as shown in Table \ref{table3}. In this table, ``$*$'' indicates that there is no requirement for the state of that qubit.

\begin{table}
	\caption{The output of rational operation}
	\begin{center}
		\begin{tabular}{cc} 
			\hline
			relational operator & $|c^{i}c^{i-1}\rangle$  \\
			\hline
			${<}$ & $|01\rangle$   \\
			${<=}$ & $|0*\rangle$ \\
			${>}$ & $|10\rangle$  \\
			${>=}$ & $|*0\rangle$ \\
			${==}$ & $|00\rangle$  \\
			${!=}$ & $|01\rangle$ or $|10\rangle$  \\
			\hline
		\end{tabular} \label{table3}
	\end{center}
\end{table}

Sometimes, instead of directly comparing two variables, the code to be tested compares the values of two expressions. Suppose the two expressions are $e_1$ and $e_2$, and their outputs are $|\varphi_1\rangle$ and $|\varphi_2\rangle$ respectively. A quantum comparator is used to compare $|\varphi_1\rangle$ and $|\varphi_2\rangle$. $|c^{i}\rangle$ and $|c^{i-1}\rangle$ record the results of the comparison, i.e., they are the flags to segment the test case space. The segmentation of the test case space by a relational operator is expressed as $U_r$:
\begin{equation}
	U_r = C\otimes e_1\otimes e_2
\end{equation}

\begin{figure}[htbp]
	\centering
	\includegraphics[scale=1]{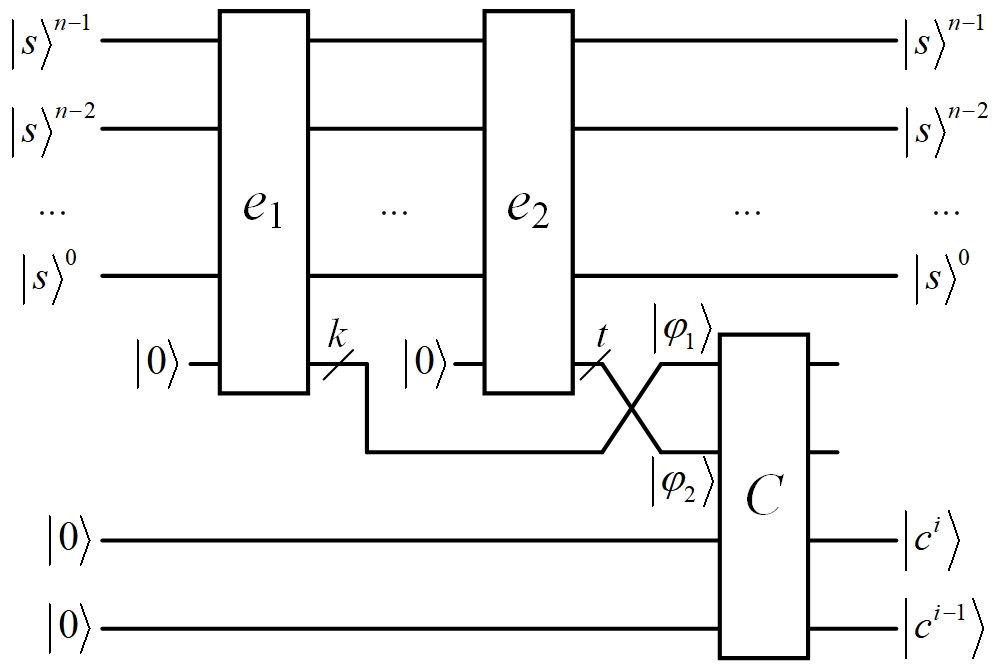}
	\caption{The segmentation of the test case space by relational operations.}
	\label{fig7}
\end{figure}

$U_r$ can segment the test case space by modifying the state of $|c^ic^{i-1}\rangle$.
\begin{equation}\label{Ur}
	\begin{split}
		&U_r(|s\rangle|0\rangle^{\otimes k}|0\rangle^{\otimes t}|00\rangle)\\
		=&C(e_1(|s\rangle|0\rangle^{\otimes k}) e_2(|s\rangle|0\rangle^{\otimes t}) |00\rangle)\\
		=&C(|s\rangle|\varphi_1\rangle|\varphi_2\rangle|00\rangle)\\
		=&|s\rangle\otimes C(|\varphi_1\rangle|\varphi_2\rangle|00\rangle)\\
		=&|s\rangle|\varphi_1\rangle|\varphi_2\rangle|c^ic^{i-1}\rangle
	\end{split}
\end{equation}
In $|s\rangle\otimes|c^ic^{i-1}\rangle$, due to the entanglement between $|s\rangle$ and $|c^ic^{i-1}\rangle$, different states of $|c^ic^{i-1}\rangle$ correspond to different subspaces of $|s\rangle$. The circuit is shown in Fig. \ref{fig7}.

In the following, we use $|c^ic^{i-1}\rangle_e$ to indicate that $|c^ic^{i-1}\rangle$ is in the output state of $e$, and $|c^ic^{i-1}\rangle_{\overline{e}}$ to indicate that $|c^ic^{i-1}\rangle$ is not in the output state of $e$, where $e=(e_1\circ e_2)$ and $\circ\in\{<,\leq,>,\geq,=,\neq\}$. For example, if $e=(e_1<e_2)$, $|c^ic^{i-1}\rangle_e=|01\rangle$, and $|c^ic^{i-1}\rangle_{\overline{e}}=|10\rangle$ or $|11\rangle$ or other non-$|01\rangle$ states.

\subsection{Logical operators} \label{logicaloperator}
\subsubsection{$T$ module}
Usually, the inputs to a logical operator are the outputs of rational operator(s). A rational operator has two outputs $|c^ic^{i-1}\rangle$. Hence, Module $T$ is defined firstly to facilitate later descriptions.

\begin{figure}[htbp]
	\centering
	\subfigure[$T_<$]{
		\includegraphics[scale=1]{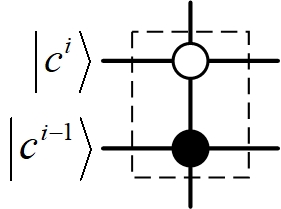}\label{figT1}
	}
	\hspace{1cm}
	\subfigure[$T_{\leq}$]{
		\includegraphics[scale=1]{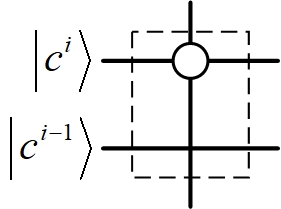}\label{figT2}
	}
	\hspace{1cm}
	\subfigure[$T_>$]{
		\includegraphics[scale=1]{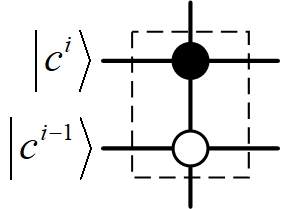}\label{figT3}
	}
	\\
	\subfigure[$T_{\geq}$]{
		\includegraphics[scale=1]{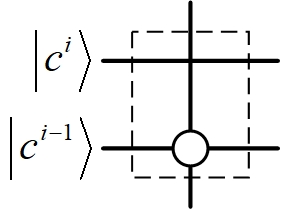}\label{figT4}
	}
	\hspace{1cm}
	\subfigure[$T_=$]{
		\includegraphics[scale=1]{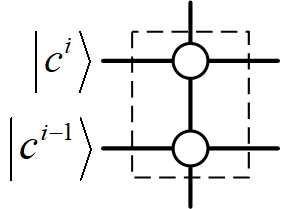}\label{figT5}
	}
	\hspace{1cm}
	\subfigure[$T_{\neq}$]{
		\includegraphics[scale=1]{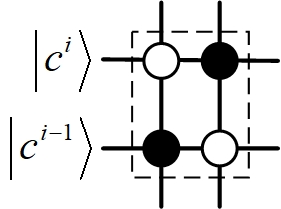}\label{figT6}
	}
	\caption{Six cases of Module $T$}
\label{fig21}
\end{figure}

$T$ is a control module that acts on two qubits $|c^ic^{i-1}\rangle$. According to Table \ref{table3},  $|c^ic^{i-1}\rangle$ have 6 states. Therefore, there are also 6 cases of $T=\{T_<,T_{\leq},T_>,T_{\geq},T_=,T_{\neq}\}$. Their circuits are shown in Fig. \ref{fig21}.

For example, in Fig. \ref{figT1}, because it is $T_<$, the state of $|c^ic^{i-1}\rangle$ is $|01\rangle$. Hence, we place a 0-control on qubit $|c^i\rangle$ and a 1-control on $|c^{i-1}\rangle$.
Thus, these two control qubits represent that the result of the previous relational operation is ``less than''.

\subsubsection{Logical operators}
There are 3 logical operators. We will give their quantum circuits one by one.

\begin{enumerate}[(1)]
	\item AND
\end{enumerate}

\begin{figure}[htbp]
	\centering
	\subfigure[The inputs are relational operations.]{
		\includegraphics[scale=1]{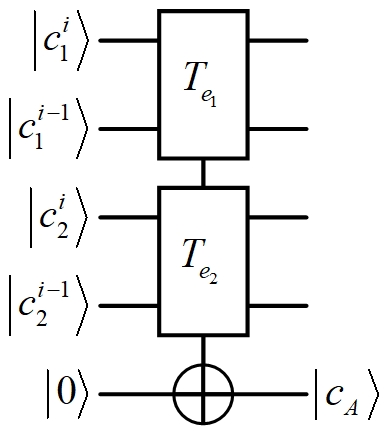}\label{fig91}
	}
	\hspace{1cm}
	\subfigure[The inputs are logical operations.]{
		\includegraphics[scale=1]{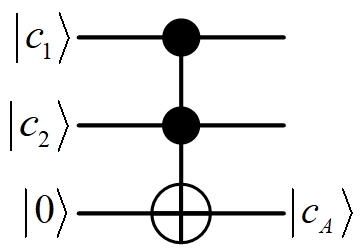}\label{fig92}
	}
	\caption{logical AND for QSE}
\label{fig9}
\end{figure}

Suppose there is an expression ${e_1\&\&e_2}$, where $e_1$ and $e_2$ are two rational operations. The logical AND in QSE is shown in Fig. \ref{fig91}, where $T_{e_1},T_{e_2}\in T$, $|c_1^i c_1^{i-1}\rangle$ are the flags of ${e_1}$, and $|c_2^i c_2^{i-1}\rangle$ are the flags of ${e_2}$. The output of logical AND is $|c_{A}\rangle$: if and only if both $e_1$ and $e_1$ are satisfied, $|c_{A}\rangle$ becomes $|1\rangle$; otherwise, it remains unchanged in $|0\rangle$ state. That is to say, $|c_{A}\rangle$ becomes a flag of logical AND.

Define
\begin{equation}
	U_{Ar} = T_{e_1}\text{-}T_{e_2}\text{-NOT}
\end{equation}
Then,
\begin{equation}\label{UAr}
	\begin{split}
		&U_{Ar}(|c_1^i c_1^{i-1}\rangle\otimes|c_2^i c_2^{i-1}\rangle\otimes|0\rangle)\\
		=&T_{e_1}\text{-}T_{e_2}\text{-NOT}(|c_1^i c_1^{i-1}\rangle\otimes|c_2^i c_2^{i-1}\rangle\otimes|0\rangle)\\
		=&|c_1^i c_1^{i-1}\rangle_{e_1}\otimes|c_2^i c_2^{i-1}\rangle_{e_2}\otimes|1\rangle+|c_1^i c_1^{i-1}\rangle_{\overline{e_1}}\otimes|c_2^i c_2^{i-1}\rangle_{{e_2}}\otimes|0\rangle\\
		&+|c_1^i c_1^{i-1}\rangle_{{e_1}}\otimes|c_2^i c_2^{i-1}\rangle_{\overline{e_2}}\otimes|0\rangle+|c_1^i c_1^{i-1}\rangle_{\overline{e_1}}\otimes|c_2^i c_2^{i-1}\rangle_{\overline{e_2}}\otimes|0\rangle
	\end{split}
\end{equation}

If $e_1$ and $e_2$ are two logical operations, it is only necessary to replace $|c_1^i c_1^{i-1}\rangle$ with $|c_1\rangle$, $|c_2^i c_2^{i-1}\rangle$ with $|c_2\rangle$, and $ T_{e_1}$ and $T_{e_2}$ with 1-control, as shown in Fig. \ref{fig92}, where $|c_1\rangle$ and $|c_2\rangle$ are the outputs of $e_1$ and $e_2$ respectively. Now
\begin{equation}
	U_{Al} = \text{CC-NOT}
\end{equation}
and
\begin{equation}\label{UAl}
	\begin{split}
		&U_{Al}(|c_1\rangle\otimes|c_2\rangle\otimes|0\rangle)\\
		=&\text{CC-NOT}(|c_1\rangle\otimes|c_2\rangle\otimes|0\rangle)\\
		=&|1\rangle\otimes|1\rangle\otimes|1\rangle+|0\rangle\otimes|1\rangle\otimes|0\rangle+|1\rangle\otimes|0\rangle\otimes|0\rangle+|0\rangle\otimes|0\rangle\otimes|0\rangle
	\end{split}
\end{equation}

\begin{enumerate}[(2)]
	\item OR
\end{enumerate}

\begin{figure}[htbp]
	\centering
	\subfigure[The inputs are relational operations.]{
		\includegraphics[scale=1]{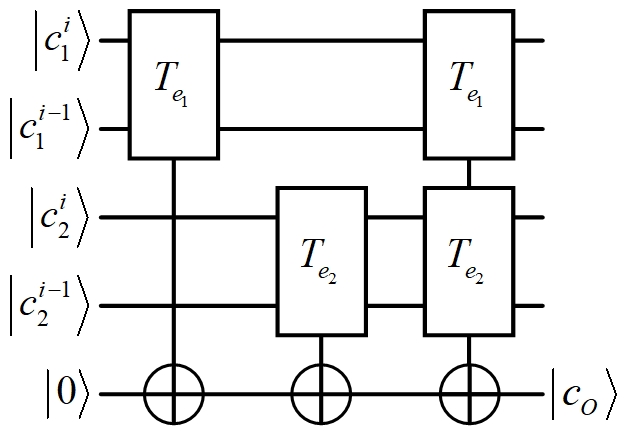}\label{fig101}
	}
	\hspace{1cm}
	\subfigure[The inputs are logical operations.]{
		\includegraphics[scale=1]{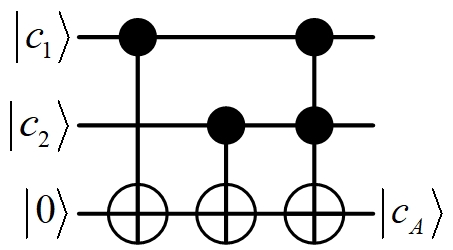}\label{fig102}
	}
	\caption{logical OR for QSE}
\label{fig10}
\end{figure}

For logical OR, there is an expression ${e_1||e_2}$. Fig. \ref{fig101} shows the logical OR in QSE if $e_1$ and $e_2$ are two rational operations. The output of logical OR is $|c_{O}\rangle$: as long as one of $e_1$ and $e_1$ is satisfied, $|c_{O}\rangle$ becomes $|1\rangle$; otherwise, it remains unchanged in $|0\rangle$ state. That is to say, $|c_{O}\rangle$ becomes a flag of logical OR.

Define
\begin{equation}
	U_{Or} =  T_{e_1}\text{-}T_{e_2}\text{-NOT}\otimes T_{e_2}\text{-NOT}\otimes T_{e_1}\text{-NOT}
\end{equation}
Then,
\begin{equation}\label{UA}
	\begin{split}
		&U_{Or}(|c_1^i c_1^{i-1}\rangle\otimes|c_2^i c_2^{i-1}\rangle\otimes|0\rangle)\\
		=&T_{e_1}\text{-}T_{e_2}\text{-NOT}\otimes T_{e_2}\text{-NOT}(T_{e_1}\text{-NOT}(|c_1^i c_1^{i-1}\rangle\otimes|c_2^i c_2^{i-1}\rangle\otimes|0\rangle))\\
		=&T_{e_1}\text{-}T_{e_2}\text{-NOT}\otimes T_{e_2}\text{-NOT}(|c_1^i c_1^{i-1}\rangle_{e_1}\otimes|c_2^i c_2^{i-1}\rangle\otimes|1\rangle+|c_1^i c_1^{i-1}\rangle_{\overline{e_1}}\otimes|c_2^i c_2^{i-1}\rangle\otimes|0\rangle)\\
		=&T_{e_1}\text{-}T_{e_2}\text{-NOT}(|c_1^i c_1^{i-1}\rangle_{e_1}\otimes|c_2^i c_2^{i-1}\rangle_{e_2}\otimes|0\rangle
		+|c_1^i c_1^{i-1}\rangle_{e_1}\otimes|c_2^i c_2^{i-1}\rangle_{\overline{e_1}}\otimes|1\rangle\\
		&+|c_1^i c_1^{i-1}\rangle_{\overline{e_1}}\otimes|c_2^i c_2^{i-1}\rangle_{e_2}\otimes|1\rangle
		+|c_1^i c_1^{i-1}\rangle_{\overline{e_1}}\otimes|c_2^i c_2^{i-1}\rangle_{\overline{e_1}}\otimes|0\rangle)\\
		=&|c_1^i c_1^{i-1}\rangle_{e_1}\otimes|c_2^i c_2^{i-1}\rangle_{e_2}\otimes|1\rangle
		+|c_1^i c_1^{i-1}\rangle_{e_1}\otimes|c_2^i c_2^{i-1}\rangle_{\overline{e_1}}\otimes|1\rangle\\
		&+|c_1^i c_1^{i-1}\rangle_{\overline{e_1}}\otimes|c_2^i c_2^{i-1}\rangle_{e_2}\otimes|1\rangle
		+|c_1^i c_1^{i-1}\rangle_{\overline{e_1}}\otimes|c_2^i c_2^{i-1}\rangle_{\overline{e_1}}\otimes|0\rangle
	\end{split}
\end{equation}

If $e_1$ and $e_2$ are two logical operations, the quantum circuit is shown in Fig. \ref{fig102} and represented as $U_{Ol}$. The migration principle is the same as in Fig. \ref{fig9} and will not be repeated.

\begin{enumerate}[(3)]
	\item NOT
\end{enumerate}

NOT does not need to be implemented with any quantum circuits. For $!e$, not matter $e$ is a rational operation or a logical operation, $e$ divides $|s\rangle$ into two subsets: one satisfies $e$ and the other does not. $!e$ just reverses the satisfiability and does not affect the division of the two subsets. Therefore, there is no need for quantum circuits to change the division of the subsets or to divide the subsets further.

\subsection{Divide the test case space}
Programs often have complex $e$ or the branch statements are nested. Therefore, multiple quantum operations are needed to be connected to continuously divide the test case space.

Define
\begin{equation}
	U_2 = U^{\otimes k}
\end{equation}
where $U\in\{U_r,U_{Ar},U_{Al},U_{Or},U_{Ol}\}$ and $k$ is a positive integer. Act $U_2$ on $|q\rangle_1$:
\begin{equation}\label{U2}
	\begin{split}
		|q\rangle_2=&U_2(|q\rangle_1)\\
		=&U^{\otimes k}(|s\rangle\otimes|0\rangle^m)\\
		=&\frac{1}{\sqrt{2}^n}\sum_{i=0}^{2^{n}-1}|s_i\rangle\otimes|c_i\rangle
	\end{split}
\end{equation}

According to the definitions of $U_r,U_{Ar},U_{Al},U_{Or},U_{Ol}$ in Section \ref{relationaloperator} and Section \ref{logicaloperator}, the qubits $|0\rangle^m$ in $|q\rangle_1$ is gradually modified based on the relational and the logical operators in the program to be tested. Eventually, through the entanglement of $|s\rangle$ and $|c\rangle$, the test case space is divided into multiple subsets. The values belonging to the same subset are test cases that can cover the same branch.

\section{Experiments}
\subsection{An example}
\subsubsection{The division of the test space}
\begin{figure}[htbp]
	\centering
	\includegraphics[scale=1]{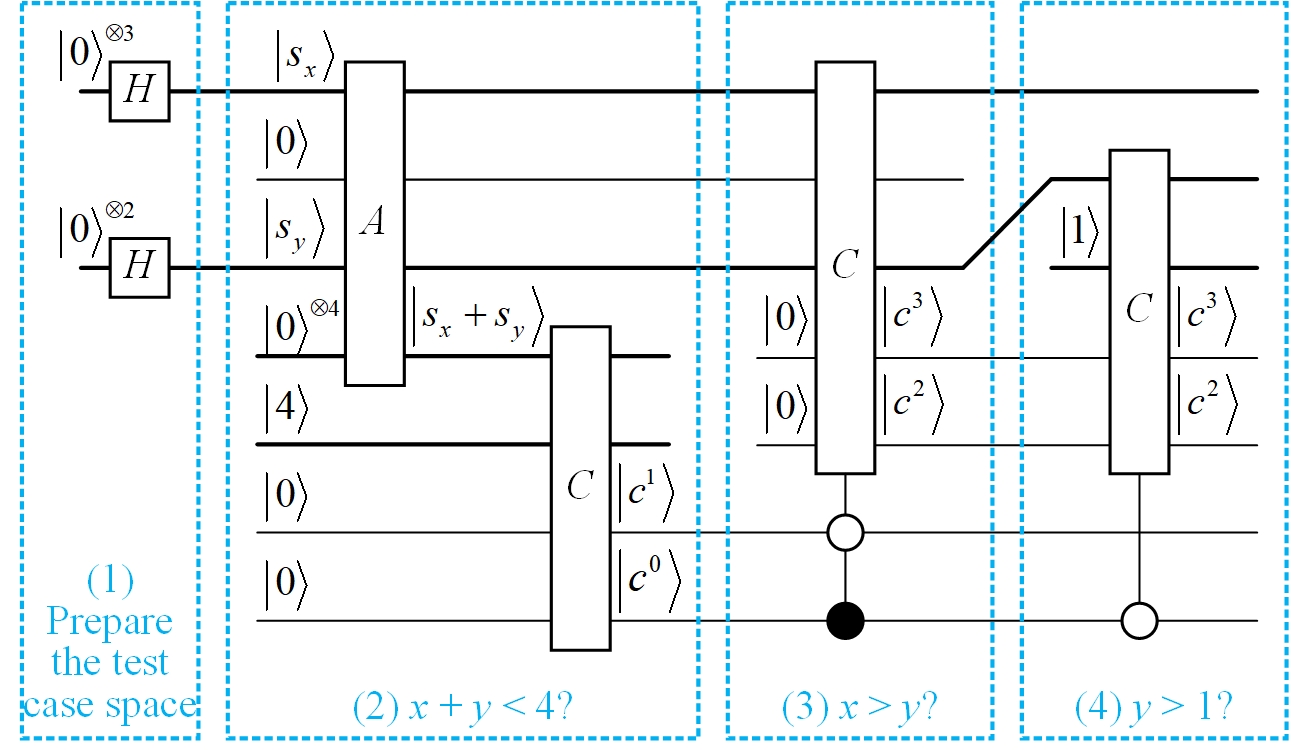}
	\caption{QSE circuit for example in Fig. \ref{fig1}}
	\label{fig13}
\end{figure}

The program shown in Fig. \ref{fig1} is used as an example to further illustrate how QSE works. There are 3 branch statements in the program. Coupled with the process of preparing the test case space, the quantum circuit consists of 4 parts as shown in Fig. \ref{fig13}.

\begin{enumerate}[(1)]
	\item Prepare the test case space
\end{enumerate}

3 and 2 qubits are used to represent variables $x$ and $y$ respectively. Hence, 5 $H$ quantum gates   transform the initial state $|0\rangle^{\otimes 5}$ to state $|s_x\rangle\otimes|s_y\rangle$, i.e.,
\begin{equation*}
	\begin{split}
		&H^{\otimes 5}(|0\rangle^{\otimes 5})=(H|0\rangle)^{\otimes 3}\otimes (H|0\rangle)^{\otimes 2}\\
		=&\frac{1}{\sqrt{2}^3}\sum_{i=0}^{7}|i\rangle\otimes\frac{1}{\sqrt{2}^2}\sum_{i=0}^{3}|i\rangle=|s_x\rangle\otimes|s_y\rangle
	\end{split}
\end{equation*}
That is to say, $|s_x\rangle$ stores $0\sim7$ and $|s_y\rangle$ stores $0\sim3$. This is the test case space.

\begin{enumerate}[(2)]
	\item $x+y<4$?
\end{enumerate}

The outermost branch statement is to determine whether $x+y$ is less than 4. The quantum adder ``$A$'' is used to get the sum of $x$ and $y$. We add a $|0\rangle$ qubit as the highest bit of $|s_y\rangle$ to make  $|s_y\rangle$ and  $|s_x\rangle$ both have 3 qubits. The quantum comparator ``$C$'' is used to compare $|s_x+s_y\rangle$ and $|4\rangle$, and the output is $|c^1c^0\rangle$. If $x+y<4$, $|c^1c^0\rangle=|01\rangle$; otherwise, $|c^1c^0\rangle=|*0\rangle$. The whole process can be described with the following equation.
\begin{equation*}
	\begin{split}
		&(C\otimes A)(|s_x\rangle|s_y\rangle\otimes|0\rangle|4\rangle|0\rangle|0\rangle)=C(A(|s_x\rangle|s_y\rangle|0\rangle)\otimes|4\rangle|0\rangle|0\rangle)\\
		=&C((|0\rangle|0\rangle|0\rangle+|0\rangle|1\rangle|1\rangle+|0\rangle|2\rangle|2\rangle+|0\rangle|3\rangle|3\rangle\\
		&+|1\rangle|0\rangle|1\rangle+|1\rangle|1\rangle|2\rangle+|1\rangle|2\rangle|3\rangle+|1\rangle|3\rangle|4\rangle\\
		&+|2\rangle|0\rangle|2\rangle+|2\rangle|1\rangle|3\rangle+|2\rangle|2\rangle|4\rangle+|2\rangle|3\rangle|5\rangle\\
		&+|3\rangle|0\rangle|3\rangle+|3\rangle|1\rangle|4\rangle+|3\rangle|2\rangle|5\rangle+|3\rangle|3\rangle|6\rangle\\
		&+|4\rangle|0\rangle|4\rangle+|4\rangle|1\rangle|5\rangle+|4\rangle|2\rangle|6\rangle+|4\rangle|3\rangle|7\rangle\\
		&+|5\rangle|0\rangle|5\rangle+|5\rangle|1\rangle|6\rangle+|5\rangle|2\rangle|7\rangle+|5\rangle|3\rangle|8\rangle\\
		&+|6\rangle|0\rangle|6\rangle+|6\rangle|1\rangle|7\rangle+|6\rangle|2\rangle|8\rangle+|6\rangle|3\rangle|9\rangle\\
		&+|7\rangle|0\rangle|7\rangle+|7\rangle|1\rangle|8\rangle+|7\rangle|2\rangle|9\rangle+|7\rangle|3\rangle|10\rangle)\otimes|4\rangle|0\rangle|0\rangle)\\
		=&|0\rangle|0\rangle C(|0\rangle|4\rangle|0\rangle|0\rangle)+|0\rangle|1\rangle C(|1\rangle|4\rangle|0\rangle|0\rangle)+|0\rangle|2\rangle C(|2\rangle|4\rangle|0\rangle|0\rangle)\\
		&+|0\rangle|3\rangle C(|3\rangle|4\rangle|0\rangle|0\rangle)+|1\rangle|0\rangle C(|1\rangle|4\rangle|0\rangle|0\rangle)+|1\rangle|1\rangle C(|2\rangle|4\rangle|0\rangle|0\rangle)\\
		&+|1\rangle|2\rangle C(|3\rangle|4\rangle|0\rangle|0\rangle)+|1\rangle|3\rangle C(|4\rangle|4\rangle|0\rangle|0\rangle)+|2\rangle|0\rangle C(|2\rangle|4\rangle|0\rangle|0\rangle)\\
		&+|2\rangle|1\rangle C(|3\rangle|4\rangle|0\rangle|0\rangle)+|2\rangle|2\rangle C(|4\rangle|4\rangle|0\rangle|0\rangle)+|2\rangle|3\rangle C(|5\rangle|4\rangle|0\rangle|0\rangle)\\
		&+|3\rangle|0\rangle C(|3\rangle|4\rangle|0\rangle|0\rangle)+|3\rangle|1\rangle C(|4\rangle|4\rangle|0\rangle|0\rangle)+|3\rangle|2\rangle C(|5\rangle|4\rangle|0\rangle|0\rangle)\\
		&+|3\rangle|3\rangle C(|6\rangle|4\rangle|0\rangle|0\rangle)+|4\rangle|0\rangle C(|4\rangle|4\rangle|0\rangle|0\rangle)+|4\rangle|1\rangle C(|5\rangle|4\rangle|0\rangle|0\rangle)\\
		&+|4\rangle|2\rangle C(|6\rangle|4\rangle|0\rangle|0\rangle)+|4\rangle|3\rangle C(|7\rangle|4\rangle|0\rangle|0\rangle)+|5\rangle|0\rangle C(|5\rangle|4\rangle|0\rangle|0\rangle)\\
		&+|5\rangle|1\rangle C(|6\rangle|4\rangle|0\rangle|0\rangle)+|5\rangle|2\rangle C(|7\rangle|4\rangle|0\rangle|0\rangle)+|5\rangle|3\rangle C(|8\rangle|4\rangle|0\rangle|0\rangle)\\
		&+|6\rangle|0\rangle C(|6\rangle|4\rangle|0\rangle|0\rangle)+|6\rangle|1\rangle C(|7\rangle|4\rangle|0\rangle|0\rangle)+|6\rangle|2\rangle C(|8\rangle|4\rangle|0\rangle|0\rangle)\\
		&+|6\rangle|3\rangle C(|9\rangle|4\rangle|0\rangle|0\rangle)+|7\rangle|0\rangle C(|7\rangle|4\rangle|0\rangle|0\rangle)+|7\rangle|1\rangle C(|8\rangle|4\rangle|0\rangle|0\rangle)\\
		&+|7\rangle|2\rangle C(|9\rangle|4\rangle|0\rangle|0\rangle)+|7\rangle|3\rangle C(|10\rangle|4\rangle|0\rangle|0\rangle)\\
		=&|0\rangle|0\rangle |0\rangle|4\rangle|0\rangle|1\rangle+|0\rangle|1\rangle |1\rangle|4\rangle|0\rangle|1\rangle+|0\rangle|2\rangle |2\rangle|4\rangle|0\rangle|1\rangle+|0\rangle|3\rangle |3\rangle|4\rangle|0\rangle|1\rangle\\
		&+|1\rangle|0\rangle |1\rangle|4\rangle|0\rangle|1\rangle+|1\rangle|1\rangle |2\rangle|4\rangle|0\rangle|1\rangle+|1\rangle|2\rangle |3\rangle|4\rangle|0\rangle|1\rangle+|1\rangle|3\rangle |4\rangle|4\rangle|0\rangle|0\rangle\\
		&+|2\rangle|0\rangle |2\rangle|4\rangle|0\rangle|1\rangle+|2\rangle|1\rangle |3\rangle|4\rangle|0\rangle|1\rangle+|2\rangle|2\rangle |4\rangle|4\rangle|0\rangle|0\rangle+|2\rangle|3\rangle |5\rangle|4\rangle|1\rangle|0\rangle\\
		&+|3\rangle|0\rangle |3\rangle|4\rangle|0\rangle|1\rangle+|3\rangle|1\rangle |4\rangle|4\rangle|0\rangle|0\rangle+|3\rangle|2\rangle |5\rangle|4\rangle|1\rangle|0\rangle+|3\rangle|3\rangle |6\rangle|4\rangle|1\rangle|0\rangle\\
		&+|4\rangle|0\rangle |4\rangle|4\rangle|0\rangle|0\rangle+|4\rangle|1\rangle |5\rangle|4\rangle|1\rangle|0\rangle+|4\rangle|2\rangle |6\rangle|4\rangle|1\rangle|0\rangle+|4\rangle|3\rangle |7\rangle|4\rangle|1\rangle|0\rangle\\
		&+|5\rangle|0\rangle |5\rangle|4\rangle|1\rangle|0\rangle+|5\rangle|1\rangle |6\rangle|4\rangle|1\rangle|0\rangle+|5\rangle|2\rangle |7\rangle|4\rangle|1\rangle|0\rangle+|5\rangle|3\rangle |8\rangle|4\rangle|1\rangle|0\rangle\\
		&+|6\rangle|0\rangle |6\rangle|4\rangle|1\rangle|0\rangle+|6\rangle|1\rangle |7\rangle|4\rangle|1\rangle|0\rangle+|6\rangle|2\rangle |8\rangle|4\rangle|1\rangle|0\rangle+|6\rangle|3\rangle |9\rangle|4\rangle|1\rangle|0\rangle\\
		&+|7\rangle|0\rangle |7\rangle|4\rangle|1\rangle|0\rangle+|7\rangle|1\rangle |8\rangle|4\rangle|1\rangle|0\rangle+|7\rangle|2\rangle |9\rangle|4\rangle|1\rangle|0\rangle+|7\rangle|3\rangle |10\rangle|4\rangle|1\rangle|0\rangle
	\end{split}
\end{equation*}

\begin{enumerate}[(3)]
	\item $x>y$?
\end{enumerate}

If $x+y<4$, it needs to be further judged whether $x$ is greater than $y$. Hence, a $T_<$-$C$ module acts on the subspace $|s_x\rangle|s_y\rangle\otimes|c^3c^2c^1c^0\rangle$.
\begin{equation*}
	\begin{split}
		&T_<\text{-}C(|0\rangle|0\rangle |0001\rangle+|0\rangle|1\rangle |0001\rangle+|0\rangle|2\rangle |0001\rangle+|0\rangle|3\rangle |0001\rangle\\
		&+|1\rangle|0\rangle |0001\rangle+|1\rangle|1\rangle |0001\rangle+|1\rangle|2\rangle |0001\rangle+|1\rangle|3\rangle |0000\rangle\\
		&+|2\rangle|0\rangle |0001\rangle+|2\rangle|1\rangle |0001\rangle+|2\rangle|2\rangle |0000\rangle+|2\rangle|3\rangle |0010\rangle\\
		&+|3\rangle|0\rangle |0001\rangle+|3\rangle|1\rangle |0000\rangle+|3\rangle|2\rangle |0010\rangle+|3\rangle|3\rangle |0010\rangle\\
		&+|4\rangle|0\rangle |0000\rangle+|4\rangle|1\rangle |0010\rangle+|4\rangle|2\rangle |0010\rangle+|4\rangle|3\rangle |0010\rangle\\
		&+|5\rangle|0\rangle |0010\rangle+|5\rangle|1\rangle |0010\rangle+|5\rangle|2\rangle |0010\rangle+|5\rangle|3\rangle |0010\rangle\\
		&+|6\rangle|0\rangle |0010\rangle+|6\rangle|1\rangle |0010\rangle+|6\rangle|2\rangle |0010\rangle+|6\rangle|3\rangle |0010\rangle\\
		&+|7\rangle|0\rangle |0010\rangle+|7\rangle|1\rangle |0010\rangle+|7\rangle|2\rangle |0010\rangle+|7\rangle|3\rangle |0010\rangle)\\
		=&|0\rangle|0\rangle |0001\rangle+|0\rangle|1\rangle |0101\rangle+|0\rangle|2\rangle |0101\rangle+|0\rangle|3\rangle |0101\rangle\\
		&+|1\rangle|0\rangle |1001\rangle+|1\rangle|1\rangle |0001\rangle+|1\rangle|2\rangle |0101\rangle+|1\rangle|3\rangle |0000\rangle\\
		&+|2\rangle|0\rangle |1001\rangle+|2\rangle|1\rangle |1001\rangle+|2\rangle|2\rangle |0000\rangle+|2\rangle|3\rangle |0010\rangle\\
		&+|3\rangle|0\rangle |1001\rangle+|3\rangle|1\rangle |0000\rangle+|3\rangle|2\rangle |0010\rangle+|3\rangle|3\rangle |0010\rangle\\
		&+|4\rangle|0\rangle |0000\rangle+|4\rangle|1\rangle |0010\rangle+|4\rangle|2\rangle |0010\rangle+|4\rangle|3\rangle |0010\rangle\\
		&+|5\rangle|0\rangle |0010\rangle+|5\rangle|1\rangle |0010\rangle+|5\rangle|2\rangle |0010\rangle+|5\rangle|3\rangle |0010\rangle\\
		&+|6\rangle|0\rangle |0010\rangle+|6\rangle|1\rangle |0010\rangle+|6\rangle|2\rangle |0010\rangle+|6\rangle|3\rangle |0010\rangle\\
		&+|7\rangle|0\rangle |0010\rangle+|7\rangle|1\rangle |0010\rangle+|7\rangle|2\rangle |0010\rangle+|7\rangle|3\rangle |0010\rangle
	\end{split}
\end{equation*}

If and only if $|c^1c^0\rangle=|01\rangle$, $|s_x\rangle$ and $|s_y\rangle$ need to be compared, i.e., $|c^3c^2\rangle$ is changed according to $|s_x\rangle$ and $|s_y\rangle$: if $|s_x\rangle>|s_y\rangle$, $|c^3c^2\rangle=|10\rangle$; otherwise, $|c^3c^2\rangle=|0*\rangle$. As long as $|c^1c^0\rangle\neq|01\rangle$, $|c^3c^2\rangle$ remains unchanged at state $|00\rangle$.

\begin{enumerate}[(4)]
	\item $y>1$?
\end{enumerate}

If $x+y\geq4$, it needs to be further judged whether $y$ is greater than $1$. Hence, a $T_\geq$-$C$ module acts on the subspace $|s_y\rangle|1\rangle\otimes|c^3c^2c^1c^0\rangle$.
\begin{equation*}
	\begin{split}
		&T_\geq\text{-}C(|0\rangle|1\rangle |0001\rangle+|1\rangle|1\rangle |0101\rangle+|2\rangle|1\rangle |0101\rangle+|3\rangle|1\rangle |0101\rangle\\
		&+|0\rangle|1\rangle |1001\rangle+|1\rangle|1\rangle |0001\rangle+|2\rangle|1\rangle |0101\rangle+|3\rangle|1\rangle |0000\rangle\\
		&+|0\rangle|1\rangle |1001\rangle+|1\rangle|1\rangle |1001\rangle+|2\rangle|1\rangle |0000\rangle+|3\rangle|1\rangle |0010\rangle\\
		&+|0\rangle|1\rangle |1001\rangle+|1\rangle|1\rangle |0000\rangle+|2\rangle|1\rangle |0010\rangle+|3\rangle|1\rangle |0010\rangle\\
		&+|0\rangle|1\rangle |0000\rangle+|1\rangle|1\rangle |0010\rangle+|2\rangle|1\rangle |0010\rangle+|3\rangle|1\rangle |0010\rangle\\
		&+|0\rangle|1\rangle |0010\rangle+|1\rangle|1\rangle |0010\rangle+|2\rangle|1\rangle |0010\rangle+|3\rangle|1\rangle |0010\rangle\\
		&+|0\rangle|1\rangle |0010\rangle+|1\rangle|1\rangle |0010\rangle+|2\rangle|1\rangle |0010\rangle+|3\rangle|1\rangle |0010\rangle\\
		&+|0\rangle|1\rangle |0010\rangle+|1\rangle|1\rangle |0010\rangle+|2\rangle|1\rangle |0010\rangle+|3\rangle|1\rangle |0010\rangle)\\
		=&|0\rangle|1\rangle |0001\rangle+|1\rangle|1\rangle |0101\rangle+|2\rangle|1\rangle |0101\rangle+|3\rangle|1\rangle |0101\rangle\\
		&+|0\rangle|1\rangle |1001\rangle+|1\rangle|1\rangle |0001\rangle+|2\rangle|1\rangle |0101\rangle+|3\rangle|1\rangle |1000\rangle\\
		&+|0\rangle|1\rangle |1001\rangle+|1\rangle|1\rangle |1001\rangle+|2\rangle|1\rangle |1000\rangle+|3\rangle|1\rangle |1010\rangle\\
		&+|0\rangle|1\rangle |1001\rangle+|1\rangle|1\rangle |0000\rangle+|2\rangle|1\rangle |1010\rangle+|3\rangle|1\rangle |1010\rangle\\
		&+|0\rangle|1\rangle |0100\rangle+|1\rangle|1\rangle |0010\rangle+|2\rangle|1\rangle |1010\rangle+|3\rangle|1\rangle |1010\rangle\\
		&+|0\rangle|1\rangle |0110\rangle+|1\rangle|1\rangle |0010\rangle+|2\rangle|1\rangle |1010\rangle+|3\rangle|1\rangle |1010\rangle\\
		&+|0\rangle|1\rangle |0110\rangle+|1\rangle|1\rangle |0010\rangle+|2\rangle|1\rangle |1010\rangle+|3\rangle|1\rangle |1010\rangle\\
		&+|0\rangle|1\rangle |0110\rangle+|1\rangle|1\rangle |0010\rangle+|2\rangle|1\rangle |1010\rangle+|3\rangle|1\rangle |1010\rangle
	\end{split}
\end{equation*}

If and only if $|c^0\rangle=|0\rangle$, $|s_y\rangle$ and $|1\rangle$ need to be compared, i.e., $|c^3c^2\rangle$ is changed according to $|s_y\rangle$ and $|1\rangle$: if $|s_y\rangle>|1\rangle$, $|c^3c^2\rangle=|10\rangle$; otherwise, $|c^3c^2\rangle=|0*\rangle$. As long as $|c^0\rangle\neq|0\rangle$, $|c^3c^2\rangle$ remains unchanged.

Finally, the state of the subspace $|s_x\rangle|s_y\rangle\otimes|c^3c^2c^1c^0\rangle$ is
\begin{equation}\label{example}
	\begin{split}
		&{|0\rangle|0\rangle |0001\rangle}+{|0\rangle|1\rangle |0101\rangle}+{|0\rangle|2\rangle |0101\rangle}+{|0\rangle|3\rangle |0101\rangle}\\
		+&{|1\rangle|0\rangle |1001\rangle}+{|1\rangle|1\rangle |0001\rangle}+{|1\rangle|2\rangle |0101\rangle}+{|1\rangle|3\rangle |1000\rangle}\\
		+&{|2\rangle|0\rangle |1001\rangle}+{|2\rangle|1\rangle |1001\rangle}+{|2\rangle|2\rangle |1000\rangle}+{|2\rangle|3\rangle |1010\rangle}\\
		+&{|3\rangle|0\rangle |1001\rangle}+{|3\rangle|1\rangle |0000\rangle}+{|3\rangle|2\rangle |1010\rangle}+{|3\rangle|3\rangle |1010\rangle}\\
		+&{|4\rangle|0\rangle |0100\rangle}+{|4\rangle|1\rangle |0010\rangle}+{|4\rangle|2\rangle |1010\rangle}+{|4\rangle|3\rangle |1010\rangle}\\
		+&{|5\rangle|0\rangle |0110\rangle}+{|5\rangle|1\rangle |0010\rangle}+{|5\rangle|2\rangle |1010\rangle}+{|5\rangle|3\rangle |1010\rangle}\\
		+&{|6\rangle|0\rangle |0110\rangle}+{|6\rangle|1\rangle |0010\rangle}+{|6\rangle|2\rangle |1010\rangle}+{|6\rangle|3\rangle |1010\rangle}\\
		+&{|7\rangle|0\rangle |0110\rangle}+{|7\rangle|1\rangle |0010\rangle}+{|7\rangle|2\rangle |1010\rangle}+{|7\rangle|3\rangle |1010\rangle}
	\end{split}
\end{equation}

There are 4 cases of the state $|c^3c^2c^1c^0\rangle$:
\begin{itemize}
	\item {$|1001\rangle$}: $|c^1c^0\rangle=|01\rangle$ indicates $x+y<4$ and $|c^3c^2\rangle=|10\rangle$ indicates $x>y$. Hence, $|1001\rangle$ indicates $x+y<4 \ \&\&\  x>y$, which corresponds to $PC_{11}$ in classical symbolic execution.
	\item {$|0*01\rangle$}: $|c^1c^0\rangle=|01\rangle$ indicates $x+y<4$ and $|c^3c^2\rangle=|0*\rangle$ indicates $x\leq y$. Hence, $|0*01\rangle$ indicates $x+y<4 \ \&\&\  x\leq y$, which corresponds to $PC_{12}$ in classical symbolic execution.
	\item {$|10*0\rangle$}: $|c^1c^0\rangle=|*0\rangle$ indicates $x+y\geq 4$ and $|c^3c^2\rangle=|10\rangle$ indicates $y>1$. Hence, $|10*0\rangle$ indicates $x+y\geq 4 \ \&\&\  y>1$, which corresponds to $PC_{21}$ in classical symbolic execution.
	\item {$|0**0\rangle$}: $|c^1c^0\rangle=|*0\rangle$ indicates $x+y\geq 4$ and $|c^3c^2\rangle=|0*\rangle$ indicates $y\leq1$. Hence, $|0**0\rangle$ indicates $x+y\geq 4 \ \&\&\  y\leq1$, which corresponds to $PC_{22}$ in classical symbolic execution.
\end{itemize}

These 4 states of $|c^3c^2c^1c^0\rangle$ divide $|s_x\rangle|s_y\rangle$ into 4 subsets. As shown in Eq. \ref{example}, 
\begin{itemize}
	\item Subset $\{|1\rangle|0\rangle,|2\rangle|0\rangle,|3\rangle|0\rangle,|2\rangle|1\rangle\}$ contains all the test cases that can test the branch $x+y<4 \ \&\&\  x>y$.
	\item Subset $\{|0\rangle|0\rangle,|0\rangle|1\rangle,|0\rangle|2\rangle,|0\rangle|3\rangle,|1\rangle|1\rangle,|1\rangle|2\rangle\}$ contains all the test cases that can test the branch $x+y<4 \ \&\&\  x\leq y$.
	\item Subset $\{|2\rangle|2\rangle,|3\rangle|2\rangle,|4\rangle|2\rangle,|5\rangle|2\rangle,|6\rangle|2\rangle,|7\rangle|2\rangle,|1\rangle|3\rangle,|2\rangle|3\rangle,|3\rangle|3\rangle,|4\rangle|3\rangle, \\
	|5\rangle|3\rangle,|6\rangle|3\rangle,|7\rangle|3\rangle\}$ contains all the test cases that can test the branch $x+y\geq 4 \ \&\&\  y>1$.
	\item Subset $\{|4\rangle|0\rangle,|5\rangle|0\rangle,|6\rangle|0\rangle,|7\rangle|0\rangle,|3\rangle|1\rangle,|4\rangle|1\rangle,|5\rangle|1\rangle,|6\rangle|1\rangle,|7\rangle|1\rangle\}$ contains all the test cases that can test the branch $x+y\geq 4 \ \&\&\  y\leq1$.
\end{itemize}

\subsubsection{Running on a quantum computer}\label{sec412}
\begin{figure}[htbp]
	\centering
	\includegraphics[scale=0.15]{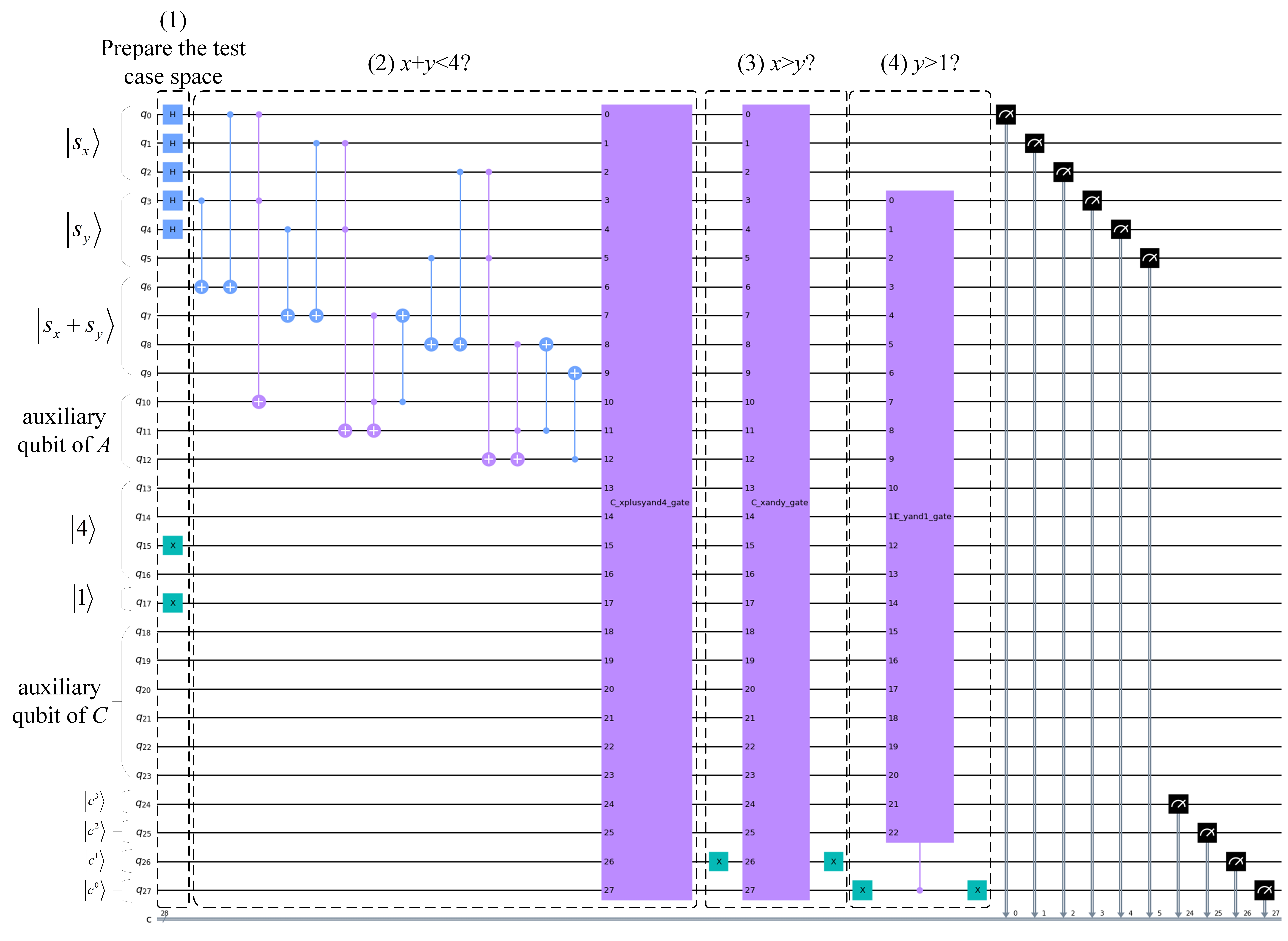}
	\caption{circuit implementation of QSE}
	\label{fig14}
\end{figure}

We use the ${ibmq \_ qasm \_ simulator}$ quantum computer on the ${IBM \; Quantum}$ platform to perform the example. The circuit is shown in Fig. \ref{fig14}. This experiment uses 28 qubits, with $q_0$ as the lowest bit and $q_{25}$ as the highest bit:
\begin{itemize}
	\item $q_2q_1q_0$ represent $|s_x\rangle$;
	\item $q_5q_4q_3$ represent $|s_y\rangle$;
	\item $q_9q_8q_7q_6$ represent $|s_x+s_y\rangle$;
	\item $q_{12}q_{11}q_{10}$ are the auxiliary qubits of the quantum adder ``$A$'';
	\item $q_{16}q_{15}q_{14}q_{13}$ are used to represent constant $\left|4\right\rangle$ and $q_{17}$ is used to represent constant $\left|1\right\rangle$;
	\item $q_{23}q_{22}q_{21}q_{20}q_{19}q_{18}$ are the auxiliary qubits of the quantum comparator ``$C$'';
	\item $q_{24}q_{25}$ are the flags $|c^3c^2\rangle$ and $q_{26}q_{27}$ are the flags $|c^1c^0\rangle$.
\end{itemize}

\begin{figure}[htbp]
	\centering
	\includegraphics[scale=0.35]{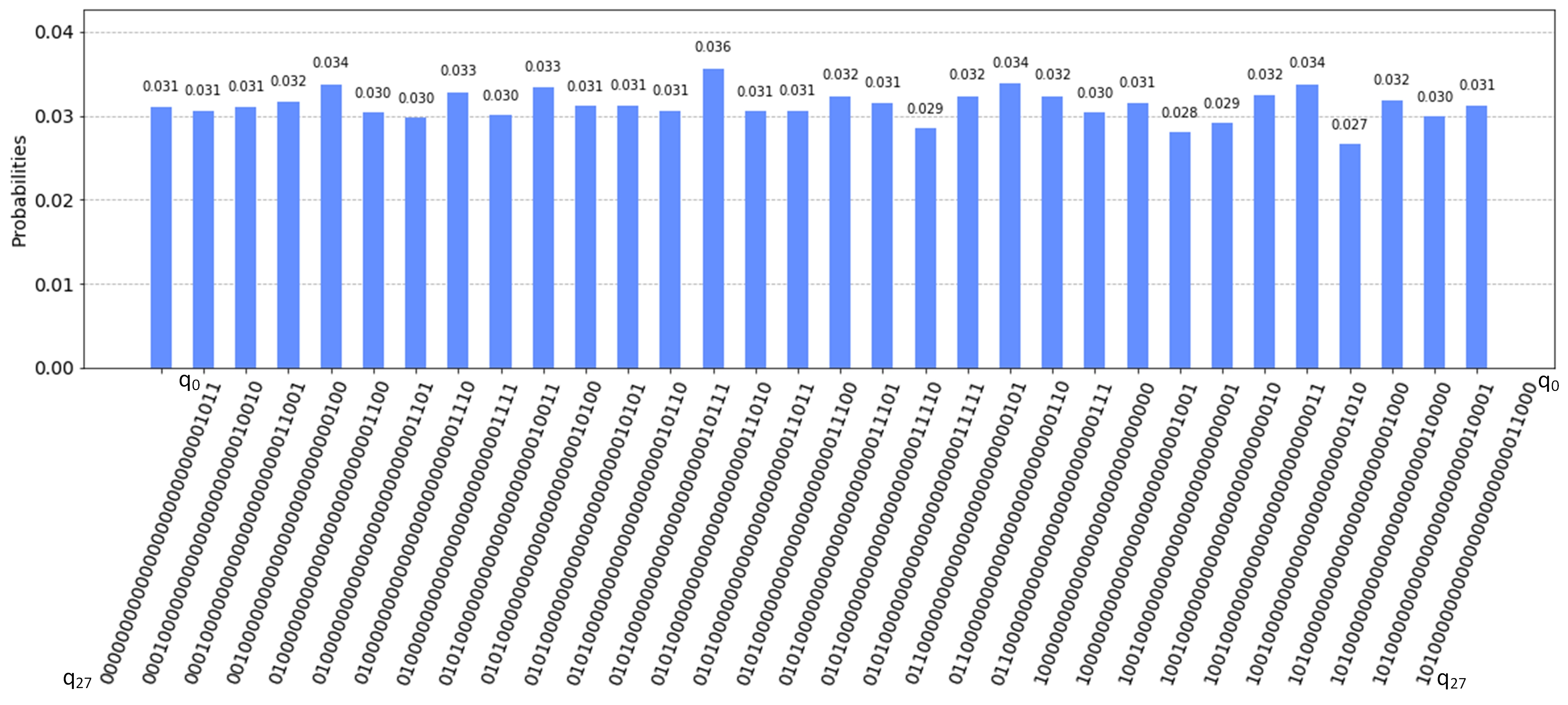}
	\caption{measurement results for the circuit in Fig. \ref{fig14}}
	\label{fig15}
\end{figure}

The three purple bars in the figure are three quantum comparators. At the end of the circuit, $q_0q_1q_2q_3q_4q_5$ and $q_{24}q_{25}q_{26}q_{27}$ are measured and they have 32 results as shown in Fig. \ref{fig15}. The abscissa displays all the results and the default state of qubits that are not measured is 0. The ordinate represents the probability of each state in a total of 8192 measurements.

\begin{figure}[htbp]
	\centering
	\subfigure[$x+y<4 \ \&\&\  x>y$]
	{ \includegraphics[scale=0.3]{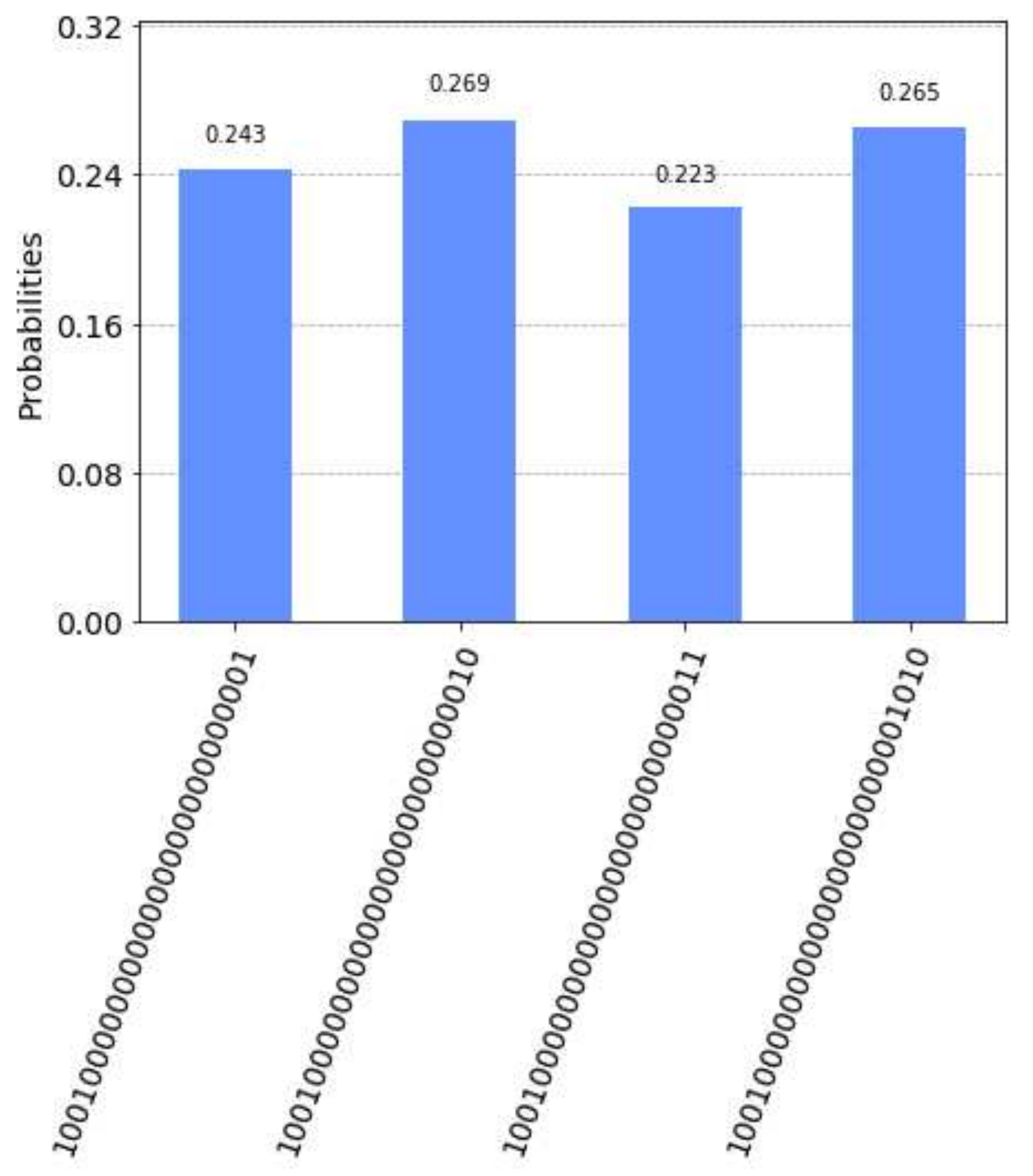}}
	\subfigure[$x+y<4 \ \&\&\  x\leq y$]
	{ \includegraphics[scale=0.3]{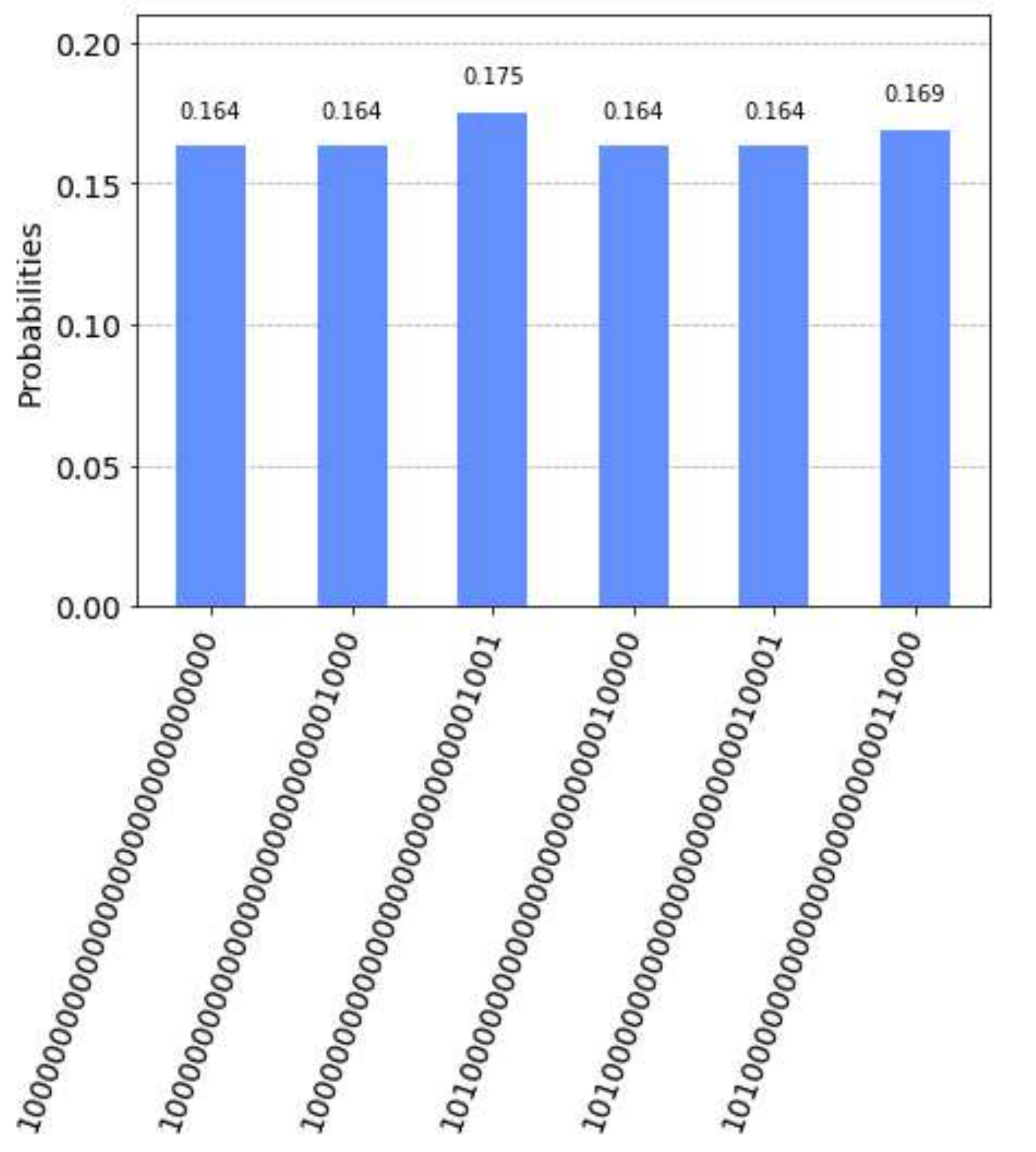}}
	\subfigure[$x+y \geq 4 \ \&\&\  y>1$]
	{ \includegraphics[scale=0.3]{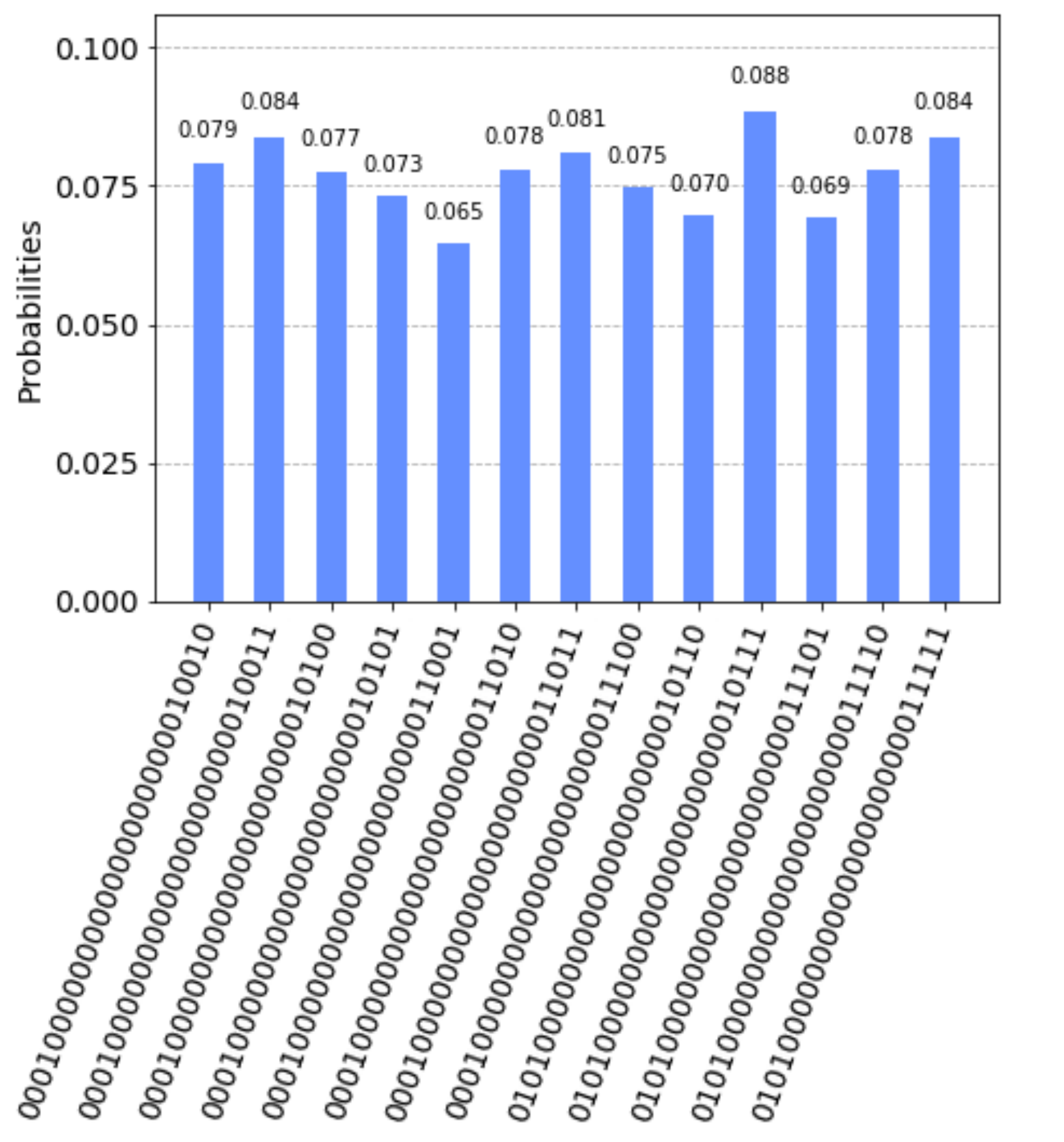}}
	\subfigure[$x+y \geq 4 \ \&\&\  y \leq 1$]
	{ \includegraphics[scale=0.3]{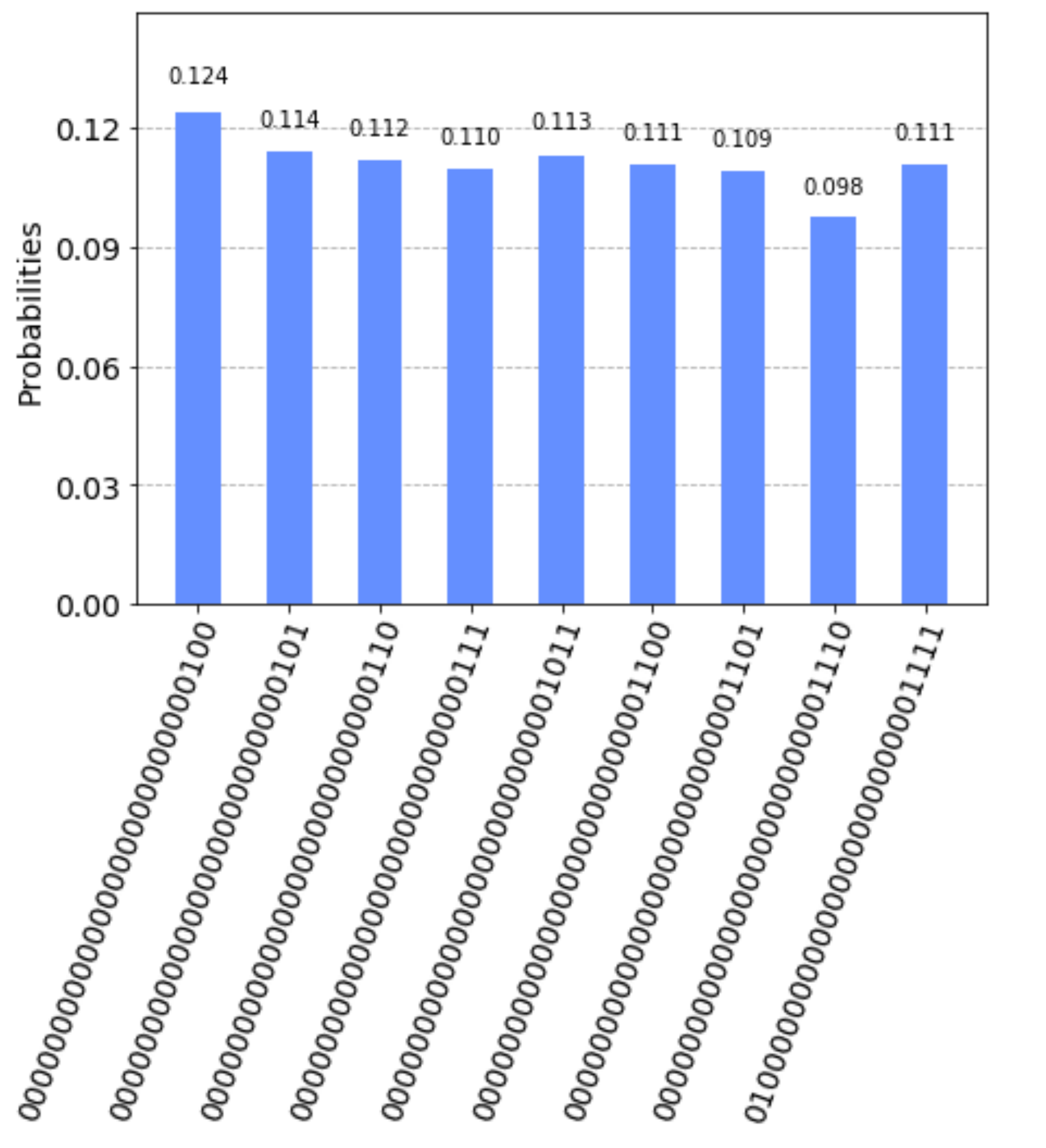}}
	\caption{Four test case spaces}
	\label{fig30}
\end{figure}

The 32 results can be divided into four test case spaces. Fig. \ref{fig30}(a) gives the measurement results whose $|c_3c_2c_1c_0\rangle=|1001\rangle$, i.e., $x+y<4 \ \&\&\  x>y$.  Fig. \ref{fig30}(b) gives the measurement results whose $|c_3c_2c_1c_0\rangle=|0*01\rangle$, i.e., $x+y<4 \ \&\&\  x\leq y$.  Fig. \ref{fig30}(c) gives the measurement results whose $|c_3c_2c_1c_0\rangle=|10*0\rangle$, i.e., $x+y \geq 4 \ \&\&\  y>1$. Fig. \ref{fig30}(d) gives the measurement results whose $|c_3c_2c_1c_0\rangle=|0**0\rangle$, i.e., $x+y \geq 4 \ \&\&\  y \leq 1$.

\subsection{Experiment data}

8 real programs are used to evaluate the performance of QSE. They come from 2 references: \cite{2014Solving} and \cite{2018VulDeePecker} as shown in Table \ref{table4}. The ``Operations'' column describes the type of operations appearing in the path conditions. The ``Line of code'' column lists the number of source code lines in the program, excluding comments and empty lines.

\begin{center}
	\begin{table}
		\centering
		\caption{Programs for the experiments}
		\begin{tabular}{cccc}
			\hline     
			Program & Operations & Line of code & From \\ 
			\hline
			dart & Polynomials & 11 & \cite{2014Solving}\\
			power & Exponential function & 20 & \cite{2014Solving}\\
			stat & Mean and std. dev. computation & 62 & \cite{2014Solving}\\
			tcas & Constant equality checks & 82 & \cite{2014Solving}\\
			early & Polynomials & 14 & \cite{2014Solving}\\
			basic00181 & Constant equality checks & 30 & \cite{2018VulDeePecker}\\
			snp3-ok & Constant equality checks & 24 & \cite{2018VulDeePecker}\\
			CWE789 & Integer computation & 141 & \cite{2018VulDeePecker}\\
			\hline     
		\end{tabular}\label{table4}
	\end{table}
\end{center}

Firstly we compare the complexity and the time consumption of CSE and QSE. The comparison results are shown in Table \ref{table5}. 
The main factor that affects the complexity of CSE are the number of path constraints.  
The main factor that affects the complexity of QSE is the number of subspace divisions. Table \ref{table5} shows that the complexity of QSE is less than that of CSE. We also compare the actual time consumption of CSE and QSE. The tool to realize CSE is JDart \cite{2016JDart}, which supports the z3 constraint solver \cite{2012Solving}. In most cases, the time consumption of QSE is also smaller than that of CSE.

\begin{center}
	\begin{table}
		\centering
		\caption{The comparison of complexity and time consumption of CSE and QSE.}
		\begin{tabular}{ccccc}\hline
			\multicolumn{1}{c}{\multirow{3}{*}{Program}} & \multicolumn{2}{c}{CSE}  & \multicolumn{2}{c}{QSE} \\ 
			\cline{2-5}
			& number of & \multicolumn{1}{c}{\multirow{2}{*}{time/s}} & number of & \multicolumn{1}{c}{\multirow{2}{*}{time/s}} \\
			& path constraints & & subspace divisions & \\
			\hline
			dart & 4 & 0.48 & 3 & 0.45 \\
			power & 11 & 1.32 & 7 & 1.05 \\
			stat & 3 & 0.36 & 2 & 0.3 \\
			tcas & 5 & 0.6 & 4 & 0.6 \\
			early & 2 & 0.24 & 1 & 0.15 \\
			basic00181 & 3 & 0.36 & 2 & 0.3 \\
			snp3-ok & 1 & 0.12 & 1 & 0.15 \\
			CWE789 & 6 & 0.72 & 3 & 0.45 \\
			\hline
		\end{tabular}\label{table5}
	\end{table}
\end{center}

\begin{figure}[htbp]
	\centering
	\includegraphics[scale=0.7]{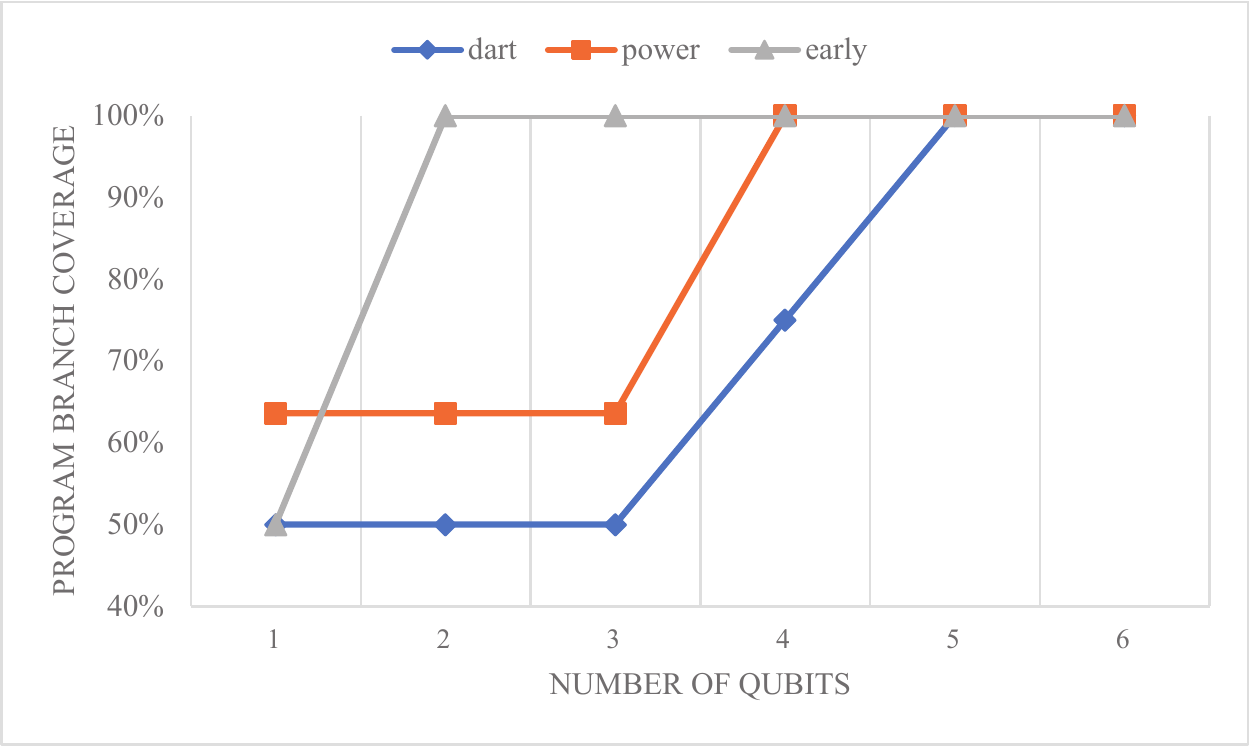}
	\caption{The relationship between the number of qubits and branch coverage}
	\label{fig35}
\end{figure}

We also show the impact of test case space on program branch coverage. In the example given in Section \ref{sec412}, three qubits are used for each variable. In fact, more or fewer qubits can affect the performance of QSE. Too few qubits make it impossible for QSE to cover all branches. Consider the more extreme case: there are 4 branches in the program, but only 1 qubit is used to store variables, i.e., there are only 2 test cases in the test case space. Such a test case space is unlikely to cover all branches. Isn't the more qubits used, the better? No. Too many qubits will increase the difficulty of QSE, and lead to the waste of quantum resources. Therefore, the smallest number of qubits that can cover all branches is the best choice. Fig \ref{fig35} shows the relationship between the number of qubits used by variables in the three programs in Table \ref{table5} and the program branch coverage. The best numbers of qubits for the three programs are 2, 4 and 5 respectively.

\section{Conclusion}
This paper proposes a quantum symbolic execution for the first time to generate high-coverage test cases. It is completely different from not only classical symbolic executions, but also quantum debugging schemes. QSE divides the test case space into subsets according to the conditional statements in the debugged program, and a subset contains all test cases that can test the same program branch. QSE not only provides a possible way to debug quantum programs, but also avoids the difficult problem of solving constraints in classical symbolic execution, which obviously reduces the difficulty and improves the efficiency of the work.\\

\noindent\textbf{Funding} This work is supported by the National Natural Science Foundation of China under Grants No.61502016.

\noindent\textbf{Data availability} All data generated or analysed during this study are included in this article.

\bibliography{ref}

\end{document}